\documentclass[11pt,a4paper]{article}

\pdfoutput=1
\usepackage[usenames,dvipsnames]{color}
\usepackage{jcappub}
\usepackage{multicol}
\usepackage{amssymb}
\usepackage{amsmath}
\usepackage{graphicx}
\usepackage{afterpage}
\usepackage{caption}
\usepackage{subcaption}
\usepackage{rotating}
\usepackage{bm}

\newcommand{\params}{\pmb{\vec{\xi}}}
\newcommand{\Eobsi}{N^{\rm c}_i}
\newcommand{\phiobsi}{\phi'_i}
\newcommand{\Etruei}{E}
\newcommand{\phitruei}{\phi}
\newcommand{\Emui}{E_{\mu}}
\newcommand{\phimui}{\phi_{\mu}}
\newcommand{\Emubari}{E_{\bar\mu}}
\newcommand{\phimubari}{\phi_{\bar\mu}}
\newcommand{\Like}{\mathcal{L}}
\newcommand{\ntot}{{n_\mathrm{tot}}}
\newcommand{\diff}{\mathrm{d}}
\newcommand{\ICfilesloc}{\href{http://icecube.wisc.edu/science/data/IC79_solarWIMP_data_release}{http://icecube.wisc.edu/science/data/IC79\_solarWIMP\_data\_release}}

\def\urltilda{\kern -.15em\lower .7ex\hbox{\~{}}\kern .04em}
\def\deg{^{\circ}}

\newcommand{\sv}{$\langle\sigma v\rangle_0=3\times10^{-26}$\,cm$^3$\,s$^{-1}$}
\title{Improved limits on dark matter annihilation in the Sun with the 79-string IceCube detector and implications for supersymmetry}
\collaborationImg{\includegraphics[width=1.5cm]{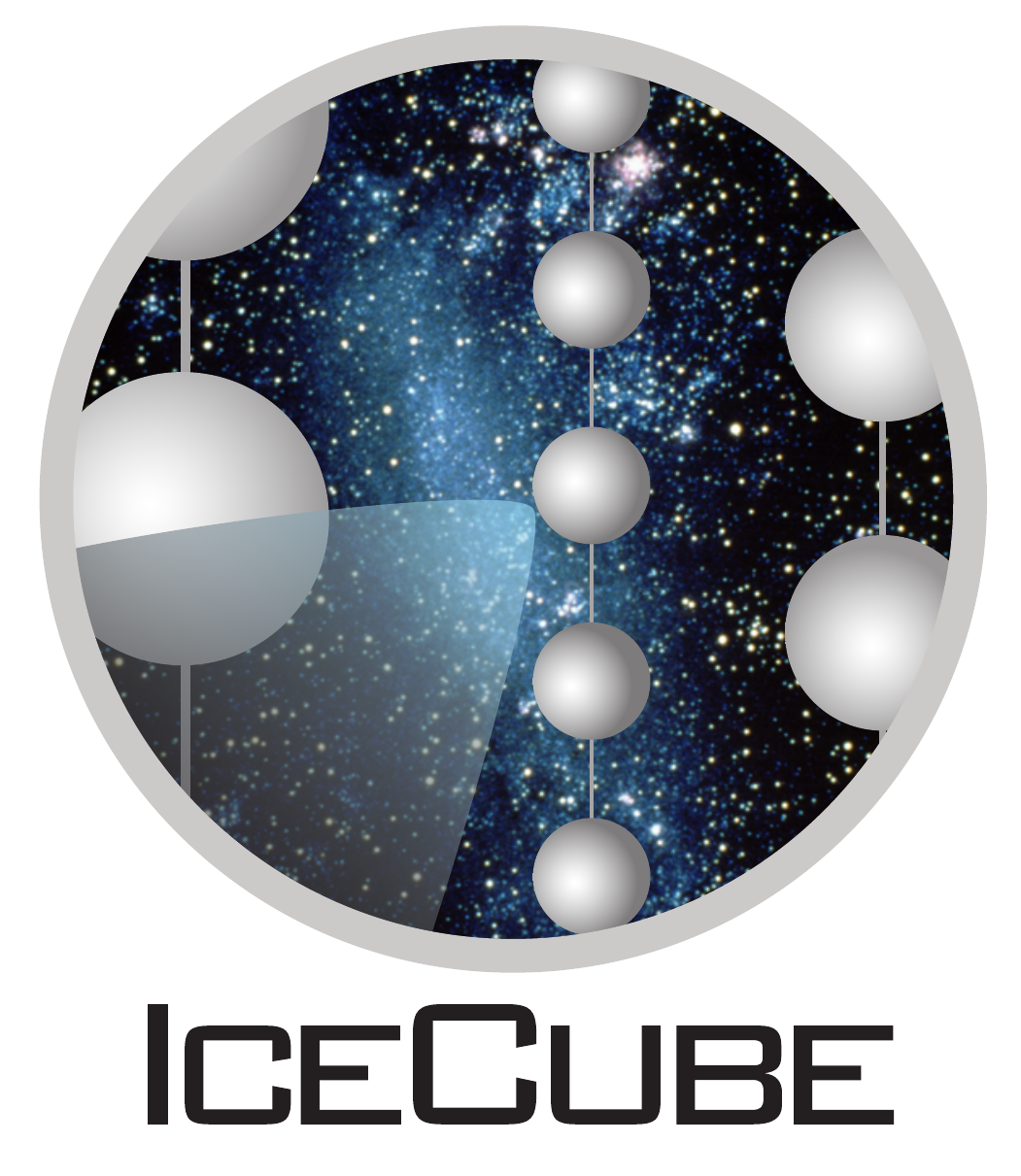}}
\collaboration{The IceCube Collaboration}
\author[2]{M.~G.~Aartsen,}
\author[33]{K.~Abraham,}
\author[50]{M.~Ackermann,}
\author[16]{J.~Adams,}
\author[12]{J.~A.~Aguilar,}
\author[30]{M.~Ahlers,}
\author[40]{M.~Ahrens,}
\author[24]{D.~Altmann,}
\author[46]{T.~Anderson,}
\author[12]{I.~Ansseau,}
% ...
\affiliation[2]{Department of Physics, University of Adelaide, Adelaide, 5005, Australia}
% ...
\affiliation[12]{Universit\'e Libre de Bruxelles, Science Faculty CP230, B-1050 Brussels, Belgium}
% ...
\affiliation[16]{Dept.~of Physics and Astronomy, University of Canterbury, Private Bag 4800, Christchurch, New Zealand}
% ...
\affiliation[24]{Erlangen Centre for Astroparticle Physics, Friedrich-Alexander-Universit\"at Erlangen-N\"urnberg, D-91058 Erlangen, Germany}
% ...
\affiliation[30]{Dept.~of Physics and Wisconsin IceCube Particle Astrophysics Center, University of Wisconsin, Madison, WI 53706, USA}
% ...
\affiliation[33]{Technische Universit\"at M\"unchen, D-85748 Garching, Germany}
% ...
\affiliation[40]{Oskar Klein Centre and Dept.~of Physics, Stockholm University, SE-10691 Stockholm, Sweden}
% ...
\affiliation[46]{Dept.~of Physics, Pennsylvania State University, University Park, PA 16802, USA}
% ...
\affiliation[50]{DESY, D-15735 Zeuthen, Germany}
% ...
\emailAdd{p.scott@imperial.ac.uk}
\emailAdd{matthias.danninger@cern.ch}

\abstract
{
  We present an improved event-level likelihood formalism for including neutrino telescope data in global fits to new physics.  We derive limits on spin-dependent dark
  matter-proton scattering by employing the new formalism in a re-analysis of data from the 79-string IceCube search for dark matter annihilation in the Sun, including explicit energy information for each event.  The new analysis excludes a number of models in the weak-scale minimal supersymmetric standard model (MSSM) for the first time.  This work is accompanied by the public release of the 79-string IceCube data, as well as an associated computer code for applying the new likelihood to arbitrary dark matter models.
}


\begin{document}

\maketitle

\section{Introduction}
\label{intro}

Searches for high-energy neutrinos from the Sun are currently the most sensitive means of probing spin-dependent interactions between protons and most models for dark matter (DM) \cite{IC79, SuperK15}.  Most analyses take a semi-model-independent approach, assuming that capture and annihilation have reached equilibrium in the Sun, and that DM annihilates exclusively into a single final state.  These assumptions are expressly violated in many concrete models for the identity of DM, including supersymmetry \cite{Wikstrom09, Trotta09, Ellis09, Ellis11, IC22Methods}.  Resulting limits are often difficult to meaningfully connect to theoretical predictions \cite{Cotta12,SM13,Trotta08,Akrami09,Strege15,Fittinocoverage,MasterCodeMSSM10,SuperBayesGC}, in part because the necessary data and likelihood functions for recasting limits to other theories are unavailable.  The computational expense required to replicate the experimental analyses for millions of parameter combinations can also be prohibitive.  All these issues arise in some form in direct detection, collider searches and other forms of indirect detection as well \cite{Scott09c,Akrami11DD,Akrami11coverage,Strege12,LATpMSSM,ATLAS15}.
This paper provides a solution to these problems for the indirect dark matter search with neutrinos.

We previously presented a 79-string search for dark matter annihilation in the Sun (IC79; \cite{IC79}), deriving limits on single annihilation channels.  We later developed a formalism (Paper I; \cite{IC22Methods}) that allows event-level neutrino telescope data to be used to constrain DM models with mixed annihilation final states, thereby allowing IceCube searches to be properly included in global fits to theories beyond the Standard Model of particle physics (BSM).  Paper I provided methods applicable to neutrinos with high energies (50\,GeV and above) that were observed with the 22-string configuration of IceCube.
This paper (Paper II) revises this formalism to include the impact of non-negligible angles between the neutrino direction and the muon produced, extending the reach of the technique to neutrino energies as low as 10 GeV.
We then apply the formalism to IC79 data and use it to rule out some example supersymmetric models.  Compared to the original IC79 analysis \cite{IC79, DanningerThesis}, which was based solely on the observed arrival directions of events, here we also include event-level energy information and an explicit treatment of the total number of observed events within the signal region, leading to an improvement in limits at high DM masses.  Extensive references on neutrino searches for dark matter and BSM global fits can be found in Paper I.

We publicly provide the fast likelihood code (\textsf{nulike}\footnote{\href{http://nulike.hepforge.org}{http://nulike.hepforge.org}}) that implements the improved analysis presented in this paper, using the public IC79 event information and detector response.  \textsf{Nulike} also provides pre-computed, fully model-independent `partial likelihoods' for every event observed by IC79, making new limits quick and easy to obtain for any annihilation final state or combination thereof.  This is a distinct advantage over the standard IceCube analysis pathway, where full signal propagation and detector simulations are required for each model.  While the approach in this paper relies on many results of the direct simulation method, such as effective areas and volumes, it provides a complete framework in which they can then be applied to essentially \textit{any} neutrino annihilation signal that can be safely treated as a point source.  The methods and the corresponding code are agnostic with respect to the details of the experiment and can be used to perform similar analyses for other neutrino telescopes, given appropriate input data in the form of event and detector response files.

In Section 2, we will provide details of the IC79 data that we use in the updated analysis, before describing the improved likelihood formalism in Section\ \ref{likelihood}.  We then show the impacts of the new analysis on generic weakly-interacting massive particle (WIMP) models in Section\ \ref{results_wimp} and models in the minimal supersymmetric standard model (MSSM) in Section\ \ref{results_mssm}. We will conclude in Section\ \ref{conclusions}.

\section{The 79-string IceCube search for dark matter}
\label{data}

\subsection{The IceCube detector}
\label{icdescription}

Completed in December 2010, the IceCube neutrino observatory~\cite{IceCube_review2} is a neutrino telescope situated at the South Pole. IceCube is installed in the glacial ice at depths of between 1450\,m and 2450\,m, instrumenting a total volume of one cubic kilometre. Digital Optical Modules (DOMs) arranged on vertical strings deep in the ice sheet record the Cherenkov light induced by relativistic charged particles, including those created by neutrinos interacting with the ice. The detection of photon yields and arrival times in DOMs allows for the reconstruction of the directions and energies of the secondaries.
In its 79-string configuration, 73 strings have a horizontal spacing of 125\,m and a vertical spacing of 17\,m between DOMs. The six remaining strings are located near the central string of IceCube and feature a reduced vertical spacing between DOMs of 7\,m and higher quantum efficiency photomultiplier tubes.  Along with the seven
surrounding regular strings, they form the DeepCore subarray~\cite{perfomancePaperDC}. The horizontal distance between strings in DeepCore is
less than 75\,m. The higher sensor density in clear ice provides an order of magnitude lower energy threshold of $\mathcal{O}$(10) \,GeV compared to the main IceCube array.

\subsection{Data samples}
\label{icevent}

In the analysis described in this paper, we start with pre-selected data from a search for WIMP dark matter annihilation in the Sun with the IceCube 79-string configuration~\cite{IC79}. This analysis uses 317 live-days of data taken between May 2010 and May 2011. As described in Refs.~\cite{IC79,DanningerThesis}, the DeepCore subarray is included for the first time in the analysis, lowering the energy threshold and extending the search to the austral summer (when neutrinos from the Sun pass downwards through the ice). In order to be sensitive to a wide range of potential WIMP masses, the analysis comprises three independent non-overlapping event selections. First, the full dataset is split into two seasonal streams, where September 22nd 2010 and March 22nd 2011 mark the beginning and end of the `summer' dataset. The `summer' sample (`summer low-energy' event selection, SL) is a dedicated low energy event sample that uses the surrounding IceCube strings as an instrumented muon veto in order to select neutrino-induced events that start within DeepCore. The `winter' dataset comprises two samples. The first sample (`winter high-energy' event selection, WH) has no particular track-containment requirement and aims to select upward-going muon tracks. The second sample (`winter low-energy' event selection, WL) is a low energy sample, and focuses on neutrino-induced muon tracks that start or are fully contained in DeepCore. The event selection was carried out separately for each independent sample. By design, the uncorrelated nature of the three datasets makes it straightforward to combine them in a joint likelihood.  The analysis in sections~\ref{results_wimp} and~\ref{results_mssm} uses the event-level data at final analysis level and corresponding signal simulations from~\cite{IC79} and~\cite{DanningerThesis}.

\subsection{Signal and background simulation}
\label{icsim}

Solar WIMP signals are simulated using \textsf{WIMPSim} \cite{Blennow08}, which describes the annihilation of WIMPs inside the Sun. \textsf{WIMPSim} simulates the production, interaction, oscillation and propagation of all three flavours of neutrinos from the core of the Sun to the detector.

Muons arising in single or coincident air showers as well as atmospheric neutrinos form the background to this analysis. We did not simulate these contributions, as they can be estimated by scrambling real data at the final analysis level (detailed within section~\ref{icbackground}).

\subsection{Calculation of detector efficiencies}
\label{iceffarea}

The effective volume $V_\mathrm{eff}(\Emui)$ of the detector for muon or anti-muon events produced through charged current interactions differs for each of the three event selections of Ref.\ \cite{IC79}. $V_\mathrm{eff}(\Emui)$ for the detection of muons from the Sun is a function of muon energy, averaged over the live-time of the respective event selections. It corresponds to an equivalent volume of 100\% detection efficiency, and is identical for both muons and anti-muons. We also calculated the effective area $A_{\rm eff}(E)$ for detection of muon neutrinos as a function of neutrino energy. We use $A_{\rm eff}(E)$ later to compute `bias factors', which account for selection effects in the analysis (see Section~\ref{biasFactor}). The effective areas for muon neutrinos and muon anti-neutrinos differ due to the differences in the (anti-)neutrino cross-sections with hadrons. All effective volumes and areas for the 79-string analysis are available online~\cite{filesloc}.

We specify the total systematic uncertainties related to the detector response at the 1\,$\sigma$ confidence level within each energy bin, in a manner similar to how it was done in Paper I. These uncertainties come from simulation studies, where identified sources of uncertainty, e.g.\ absolute DOM efficiency, photon propagation in ice, or calibration constants, were individually varied within reasonable ranges of their original values. Similarly, the uncertainties arising from limited simulation statistics are also given for each energy bin of $V_\mathrm{eff}$, at the 1\,$\sigma$ confidence level.  In the final analysis we combine these two errors in quadrature.

\subsection{Angular response}
\label{icangular}

The point spread function (PSF) describes the uncertainty in the reconstructed arrival direction of muons.
Closely following Paper I, the reduced (one-dimensional) PSF for the angular deviation $\Delta$ between the true arrival direction of a muon on the sky and its reconstructed direction is
\begin{equation}
\label{PSF}
P(\Delta) = \frac{\Delta}{\sigma_{\mu}^2} \, \exp\left[-\frac{\Delta^2}{2\sigma_{\mu}^2}\right]\,.
\end{equation}
We extract the parameter $\sigma_\mu$, which we refer to as the `mean angular error', directly from the one-dimensional PSF constructed from IceCube signal simulations.
As in Paper I, we determine $\sigma_\mu$ in the same energy bins that were used for calculating the detector efficiencies.
For simplicity, we neglect the curvature of the PSF on the sky, owing to the fact that for dark matter signals detected with DeepCore, the muon production angle is typically expected to be the dominant source of angular deviation.  We therefore restrict our analyses to signal regions of radii $\phi'_\mathrm{cut}$ around the solar position on the sky so as to minimise the error induced by this approximation (and the fact that we include the entire sky in a data-driven estimation of the background; cf.\ Sec.\ \ref{icbackground}).  We determined that $\phi'_\mathrm{cut}=20\deg$ provides satisfactory signal acceptance and background rejection for the WH sample, and $\phi'_\mathrm{cut}=40\deg$ is appropriate for the WL and SL datasets.

We associate angular uncertainties with real data events on an event-by-event basis, using the paraboloid method~\cite{paraboloid}. A paraboloid function is fitted to the muon track reconstruction likelihood space in the neighbourhood of the best fit. The resulting confidence ellipse on the sky is represented by the two principal axes, which correspond to the standard deviations of the likelihood function in each of two linearly-independent directions. The overall reconstructed likelihood track uncertainty, $\sigma_{\mathrm{para}}$ (the `paraboloid sigma'), is calculated as the mean in quadrature of the uncertainties along the two axes. Good track fits generally result in paraboloids that are narrow along both axes and therefore have small $\sigma_{\mathrm{para}}$ values.

\subsection{Energy estimator}
\label{icenergy}

\begin{figure}[tp]
\centering
  \begin{subfigure}[b]{1.0\textwidth}
  \centering
            \includegraphics[width=0.95\textwidth]{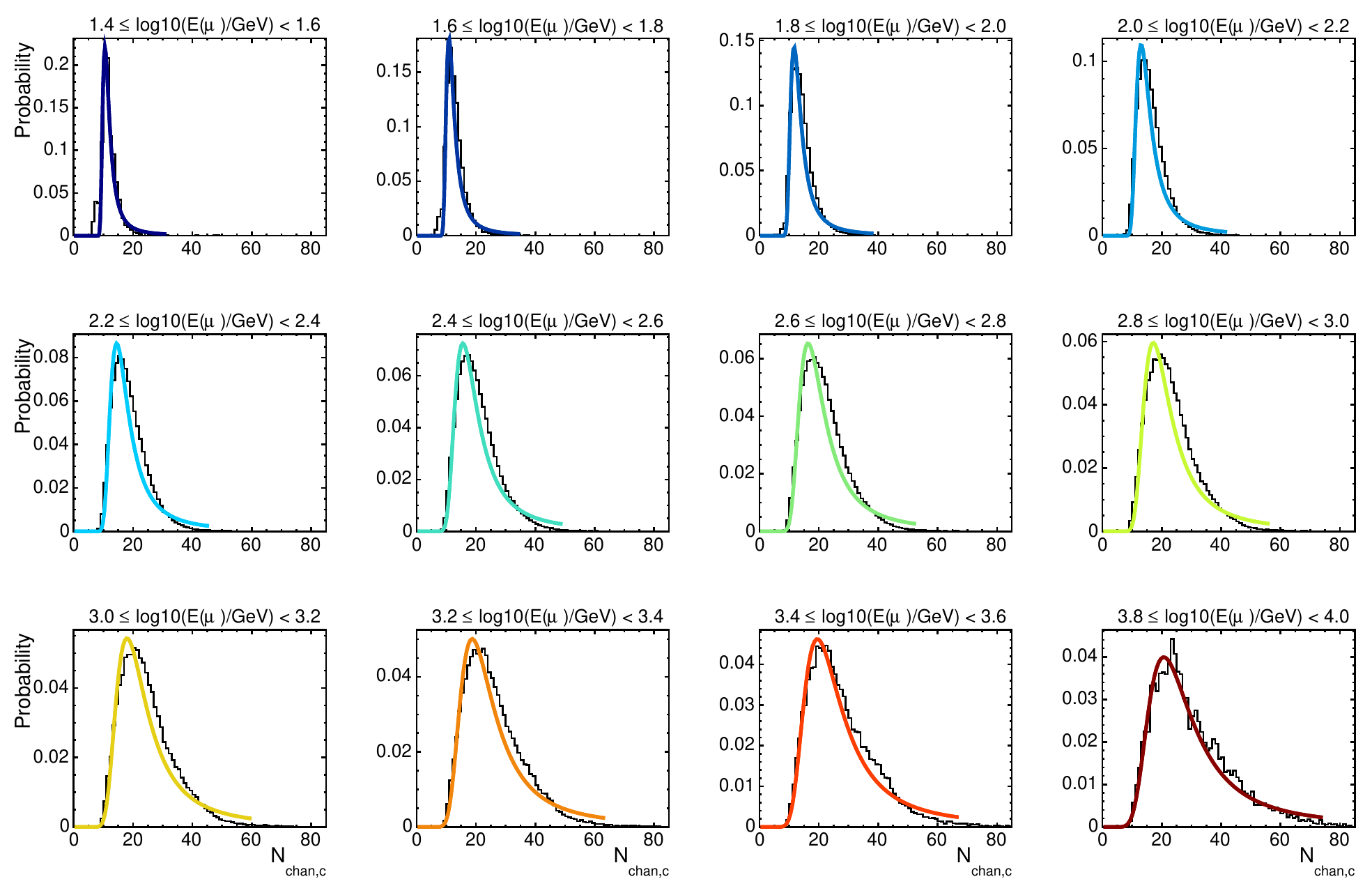}
            \caption{}
            \label{fig:energyWHa}
    \end{subfigure}
  \begin{subfigure}[b]{1.0\textwidth}
  \centering
  \includegraphics[width=0.7\textwidth]{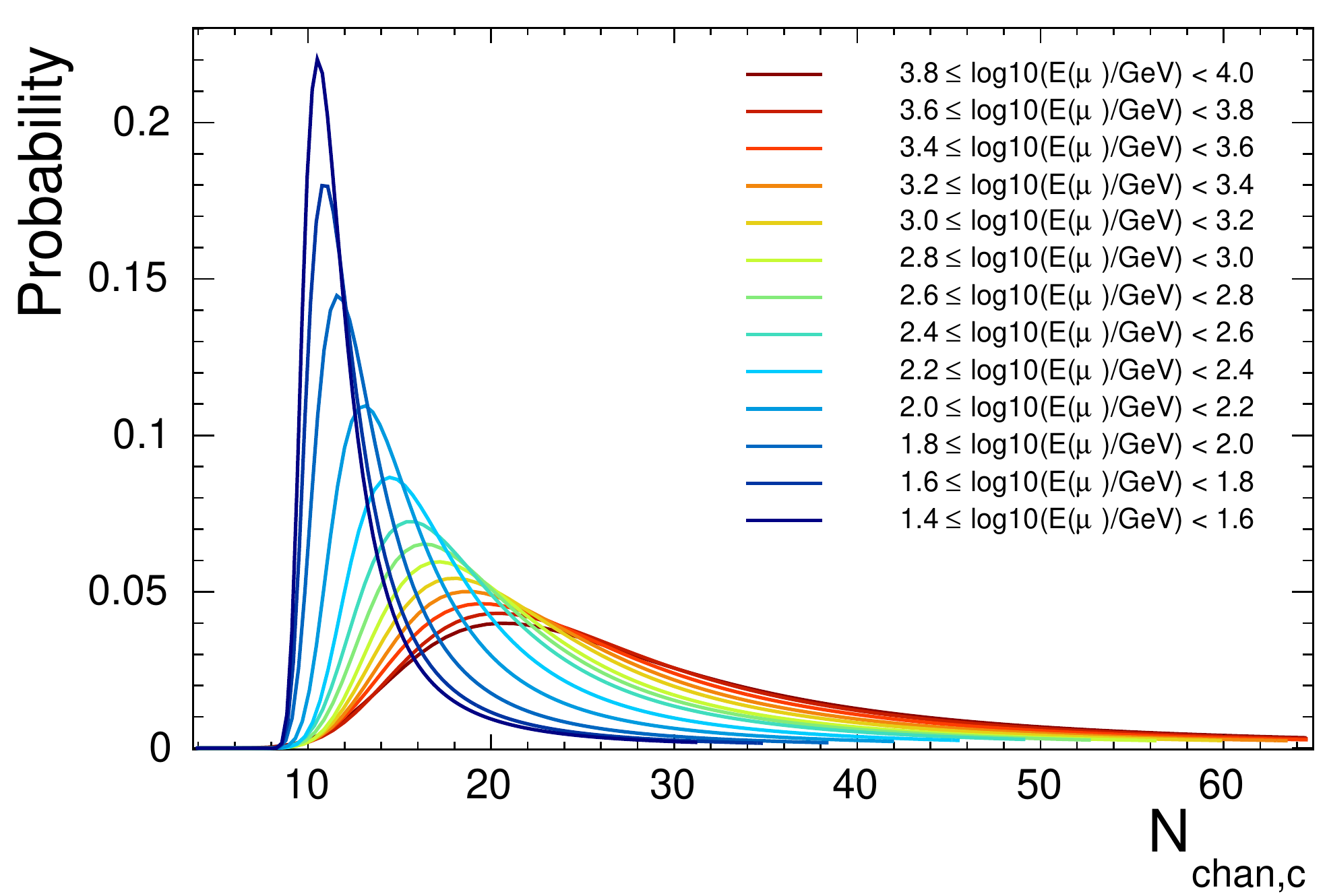}
            \caption{}
            \label{fig:energyWHb}
      \end{subfigure}
\caption{Predicted probability distributions of $N^{\rm c}_{\rm chan}$ for the WH event selection, derived from high-statistics simulations used in~\cite{IC79}. Each distribution is defined for muons having energies in a specific logarithmic energy interval of width 0.2. The fitted functions are to guide the eye only and are not used in our calculations. The lower plot compares the fitted functions, illustrating the ability to differentiate events between different energy intervals.}
\label{fig::energyWH}
\end{figure}

\begin{figure}[tp]
\centering
  \begin{subfigure}[b]{1.0\textwidth}
  \centering
            \includegraphics[width=0.95\textwidth]{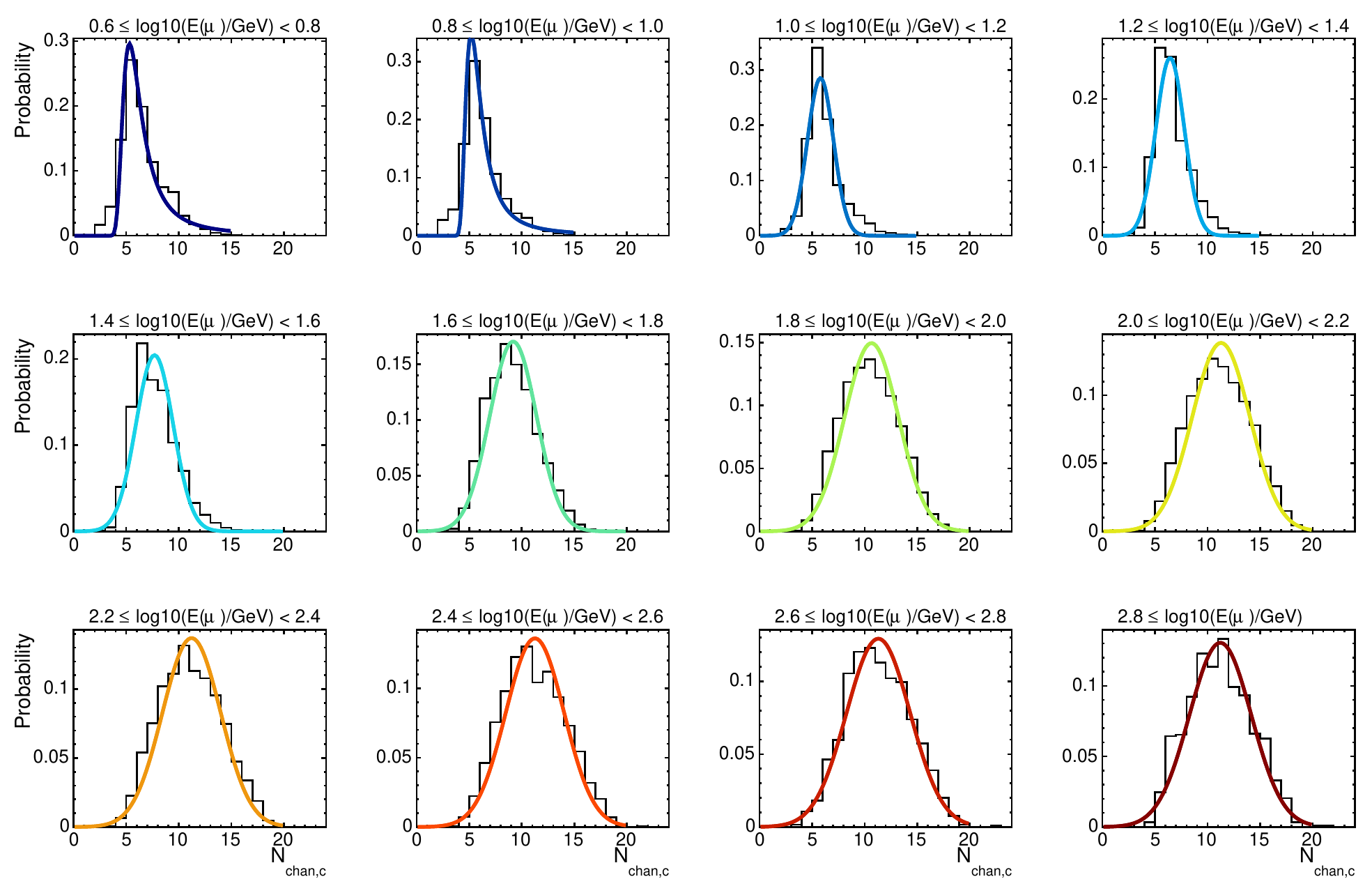}
            \caption{}
            \label{fig:energyWLa}
    \end{subfigure}
  \begin{subfigure}[b]{1.0\textwidth}
  \centering
  \includegraphics[width=0.7\textwidth]{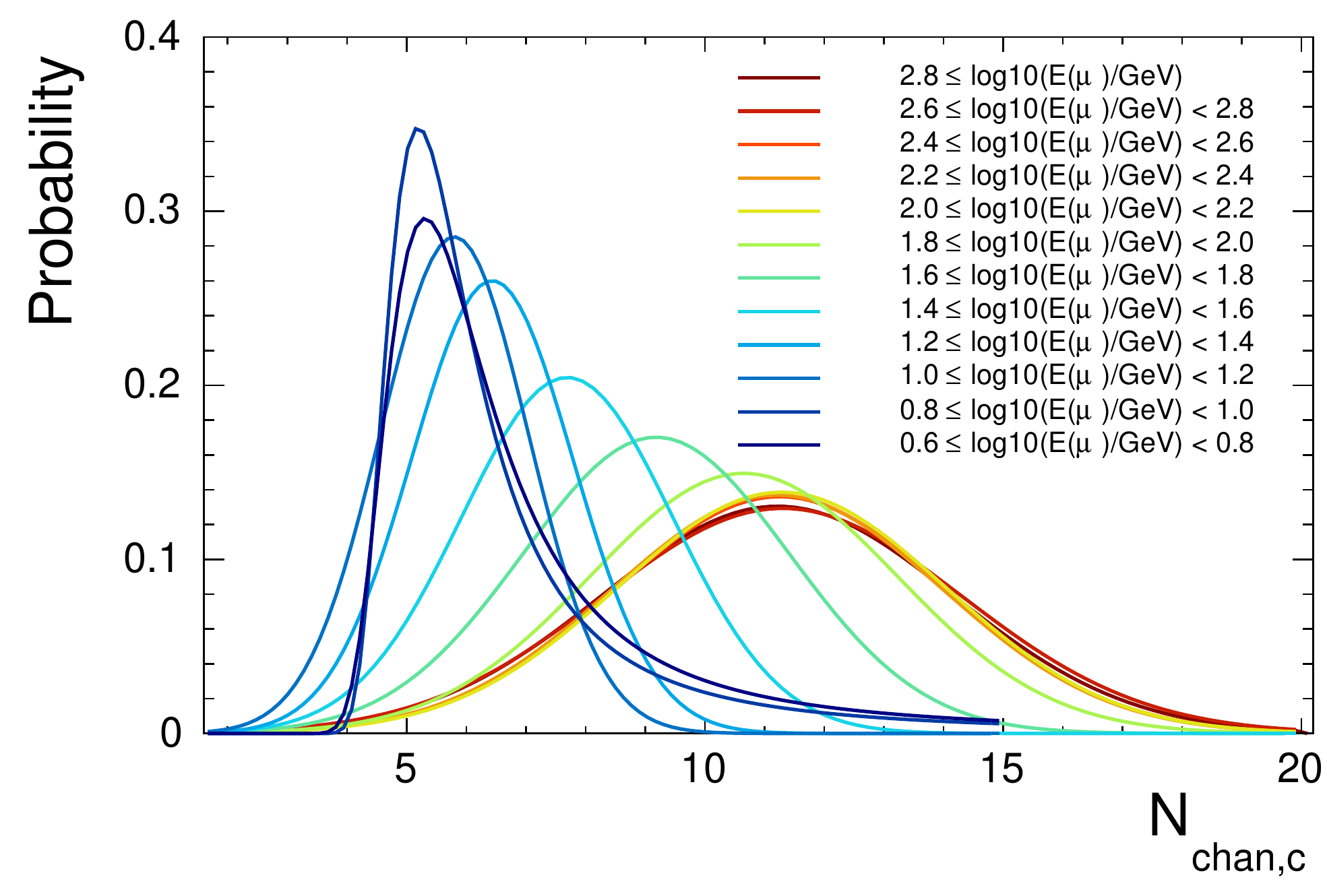}
            \caption{}
            \label{fig:energyWLb}
      \end{subfigure}
\caption{As per Fig.\ \protect\ref{fig::energyWH}, but for the WL event selection.}
\label{fig::energyWL}
\end{figure}

\begin{figure}[tp]
\centering
  \begin{subfigure}[b]{1.0\textwidth}
  \centering
            \includegraphics[width=0.95\textwidth]{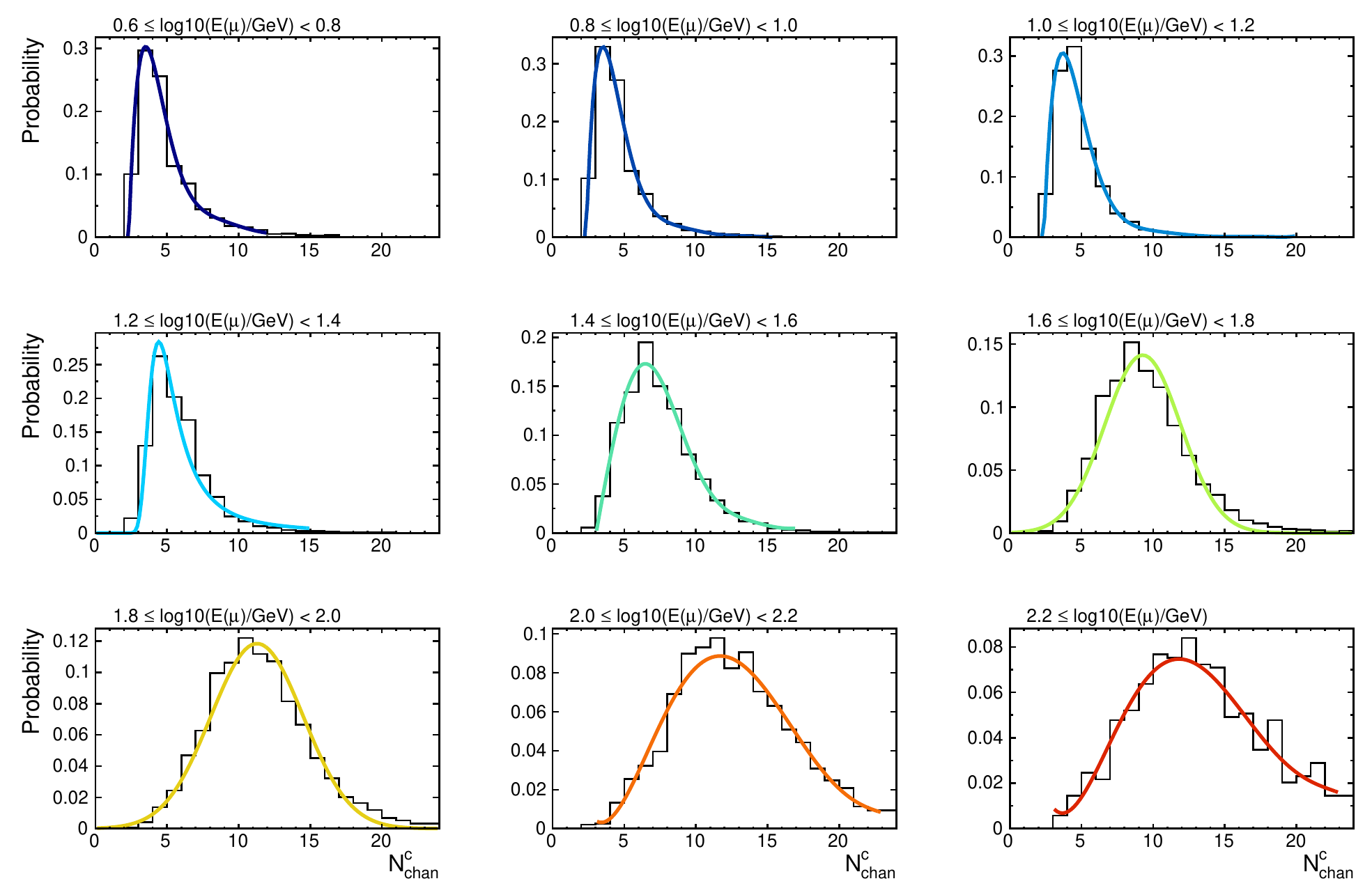}
            \caption{}
            \label{fig:energySLa}
    \end{subfigure}
  \begin{subfigure}[b]{1.0\textwidth}
  \centering
  \includegraphics[width=0.7\textwidth]{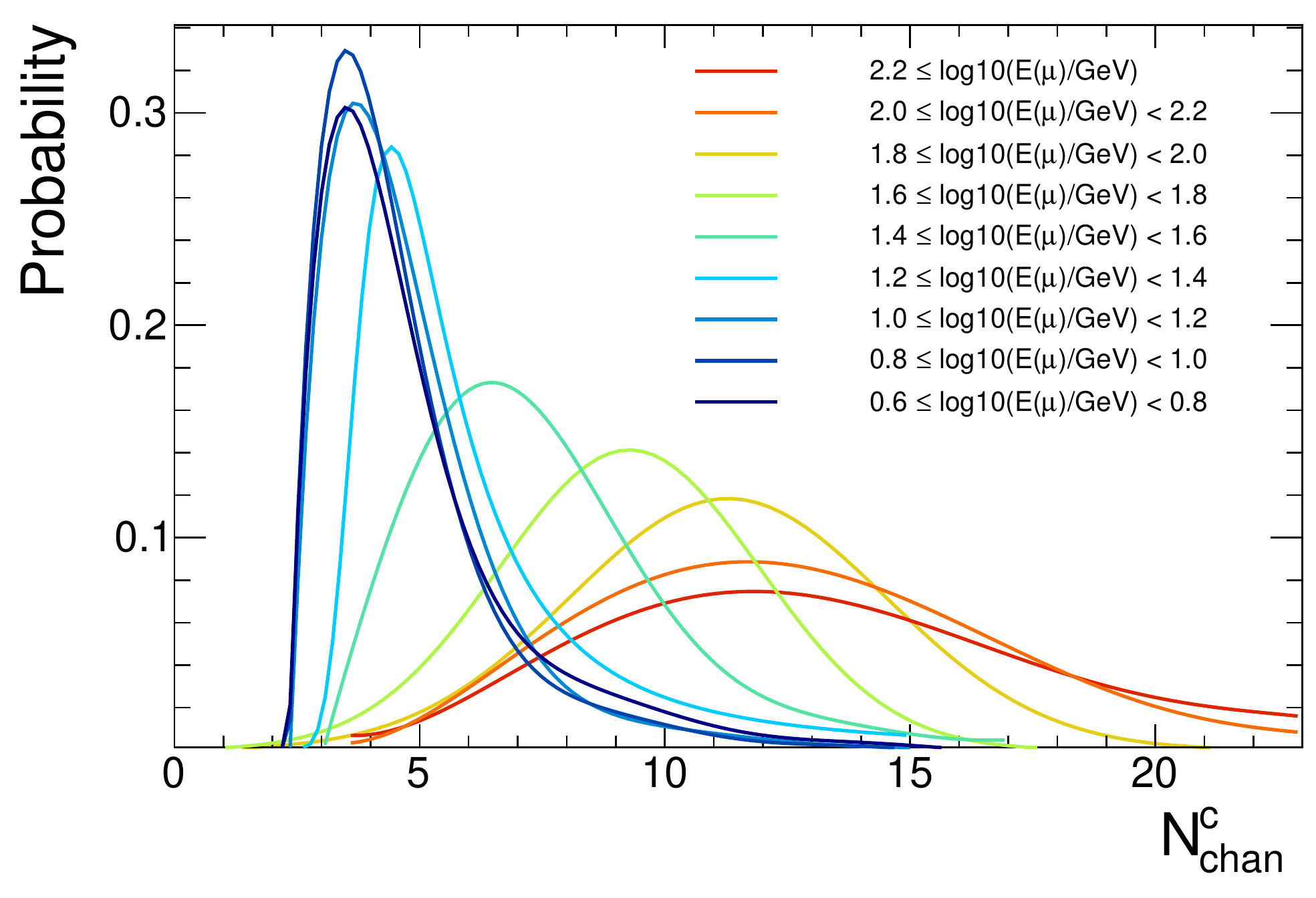}
            \caption{}
            \label{fig:energySLb}
      \end{subfigure}
\caption{As per Fig.\ \protect\ref{fig::energyWH}, but for the SL event selection.}
\label{fig::energySL}
\end{figure}

Paper I used the number of lit DOMs ($N_{\mathrm{chan}}$) as a suitable energy estimator. This definition worked well for a detector with a consistent density of optical modules, like the 22-string configuration of IceCube. This paper uses data recorded in the 79-string configuration of IceCube. This configuration includes the DeepCore subarray, which has a higher density of DOMs than the rest of the detector (Sec.~\ref{icdescription}). A simple count of lit DOMs would yield different results depending on whether the event crosses, partially crosses, or is contained within DeepCore. In an attempt to address this, we introduce a modified $N_{\mathrm{chan}}$ value, $N^{\rm c}_{\rm chan}$, which corrects for the variation in DOM density across the detector. In this context, the corrected energy proxy $N^{\rm c}_{\rm chan}$ is
\begin{equation}
\label{Nchanc}
N^{\rm c}_{\rm chan} = N_{\mathrm{chan}}^{\mathrm{IC}} + f_{\mathrm{DC}}\times N_{\mathrm{chan}}^{\mathrm{DC}}\,,
\end{equation}
where $N_{\mathrm{chan}}^{\mathrm{DC}}$ and $N_{\mathrm{chan}}^{\mathrm{IC}}$ are the number of lit DOMs in DeepCore (`standard' IceCube and `high quantum efficiency' DOMs) and the remainder of IceCube, respectively. The factor $f_{\mathrm{DC}}=0.28$ is the the ratio of the number of `standard' IceCube DOMs inside DeepCore to the total number of DOMs in DeepCore, multiplied by an additional correction factor.  The ratio accounts for the higher density of DOMs in DeepCore compared to the rest of the detector, and the additional correction factor accounts for the higher quantum efficiency of the photomultiplier tubes in the `high quantum efficiency' DOMs.

We calculated the expected distributions of observed $N^{\rm c}_{\rm chan}$ values for a series of intervals in muon energy, as we did in neutrino energy in Paper I.  Figs.\ \ref{fig::energyWH} -- \ref{fig::energySL} show these probability distributions for each event selection (WH, WL and SL) and muon energy range. The total interval in muon energy is different for each event selection due to the respective event selection criteria that are applied.
We use these probability distributions, together with the predicted energy spectrum of the signal from each WIMP model, to calculate the predicted distribution of $N^{\rm c}_{\rm chan}$.  The fitted functions in Figs.\ \ref{fig::energyWH} -- \ref{fig::energySL} are only to guide the eye; our signal predictions and likelihood calculations employ the actual distributions. The lower plots in Figs.\ \ref{fig::energyWH} -- \ref{fig::energySL} compare the fitted functions for each event selection, illustrating the ability to differentiate events between different energy intervals. 
To reach energies as low as the first interval in Figs.\ \ref{fig::energyWL} and \ref{fig::energySL}, DeepCore uses an independent, low-threshold, simple majority trigger (SMT), with a 2.5 $\mu$s time window, applied to DOMs comprising the DeepCore fiducial volume~\cite{perfomancePaperDC}. This trigger requires that three or more DOMs satisfy the so-called hard local coincidence (HLC) condition (as opposed to the threshold of eight or more DOMs, more typically used in IceCube analyses). DOMs meet the HLC condition when two or more DOMs in close proximity to each other (nearest or next-to-nearest neighbours on the same string) register hits within a 1 $\mu$s time window. This trigger is 70\% efficient for a simulated sample of atmospheric $\nu_\mu$ events of 10\,GeV neutrino energy~\cite{perfomancePaperDC}. 

More advanced energy reconstruction methods other than $N^{\rm c}_{\rm chan}$ are available in IceCube that are based on the reconstruction of charged-particle energies and topologies from the observed Cherenkov light yield~\cite{ICEnergyreco}. Here we use $N^{\rm c}_{\rm chan}$ for simplicity and robustness.

\subsection{Background estimation}
\label{icbackground}

As in Paper I, the background distributions for each event selection come directly from data.  The angular distribution of background events $\diff P_\mathrm{BG}(\phi')/\diff \phi'$ is a function of $\phi'$, the angle between the reconstructed track direction and the Sun. Muons produced in cosmic-ray showers are the dominant contributors to the background.  Their angular distribution is observed to be largely independent of azimuth, so we estimated $\diff P_\mathrm{BG}(\phi')/\diff \phi'$ from real data events at the final selection level with scrambled azimuths.  We used all observed events at the final selection level for this exercise.  Given the tight upper limit on a signal contribution in the original analysis \cite{IC79}, including the nominal signal region does not bias the background estimate.  We calculated the distribution of $N^{\rm c}_\mathrm{chan}$ due to background events, $\diff P_\mathrm{BG}(N^{\rm c}_\mathrm{chan})/\diff N^{\rm c}_\mathrm{chan}$, and observed no significant correlation between the arrival angles of events relative to the Sun and their measured $N^{\rm c}_\mathrm{chan}$ values.

\subsection{Data format, public code and availability}
\label{icdata}

Full event data from the analysis of Ref.\ \cite{IC79}, including angles, $N^{\rm c}_\mathrm{chan}$ values and paraboloid sigmas, can be found at \ICfilesloc.  Effective areas and volumes, along with $N^{\rm c}_\mathrm{chan}$ and angular responses, can be found at the same location.

The \textsf{nulike} code can be downloaded from \href{http://nulike.hepforge.org}{http://nulike.hepforge.org}. The release of \textsf{nulike} coincides with the release of \textsf{DarkSUSY v5.1.3}.  This release of \textsf{DarkSUSY} provides optimised interpolation routines for \textsf{WIMPSim} \cite{Blennow08} outputs contained in \textsf{DarkSUSY}, and ensures that they are fully compatible with the parallel likelihood routines in \textsf{nulike} (i.e.\ the routines one would use together with \textsf{nulike} are threadsafe in the latest \textsf{DarkSUSY} release).

\section{Likelihood functions}
\label{likelihood}

\subsection{General form}

The primary improvement in the likelihood treatment here compared to Paper I \cite{IC22Methods} is that we allow for differences between the arrival directions of neutrinos ($\phi$) and the muons they produce ($\phi_\mu$).  At neutrino energies above \ $\mathcal{O}$(100\,GeV), to a good approximation one can neglect the difference between $\phi$ and $\phi_\mu$.  This was the case for all data and calculations considered in Paper I.  With the DeepCore infill array however, the actual 79- and 86-string IceCube configurations are sensitive to neutrino energies even below 10\,GeV.  For example, for a neutrino of energy 10\,GeV producing a muon of 4\,GeV, $\phi - \phi_\mu$ can be as large as 30 degrees, and must therefore be explicitly included in all calculations.

The distribution of muon production angles introduces an explicit energy dependence to the detector PSF.  This improves on our earlier approximation that the detector response factorises into separate functions of angle and energy (Eq.\ 3.6 in Paper I).  In this paper we therefore work with the general form of the unbinned likelihood,
\begin{equation}
\label{unbinned_like}
\Like_\mathrm{unbin} \equiv \Like_\mathrm{num}(\ntot|\theta_\mathrm{tot})\prod_{i=1}^{\ntot} \int_0^\pi\int_0^\infty Q(\Eobsi, \phiobsi | \Etruei, \phitruei)\frac{\diff^2 P}{\diff \Etruei\,\diff \phitruei}(\Etruei,\phitruei,\params)\,\diff \Etruei\, \diff \phitruei\,.
\end{equation}
The vector $\params$ refers to the parameters of a given BSM model.  $\Eobsi$ and $\phiobsi$ are the actual observed event-level data for the $i$th event of $\ntot$ total events.  $\Eobsi$ in this analysis is the generalised $N_{\rm chan}$, whereas $\phiobsi$ is the angle between the reconstructed muon track and the direction of the Sun.  As in Paper I, $Q(\Eobsi, \phiobsi | \Etruei, \phitruei)$ is the probability density (in effective units of inverse angle and $N^{\rm c}_{\rm chan}$) for observing $\Eobsi$ and $\phiobsi$ for the $i$th event when the true values of the incoming neutrino energy and angle relative to the Sun are $\Etruei$ and $\phitruei$, respectively.

The prefactor $\Like_\mathrm{num}$ is the number likelihood for observing $\ntot$ events given a prediction $\theta_\mathrm{tot}$, marginalised over the systematic error on the predicted number of events. This is
\begin{equation}
\label{number}
\Like_\mathrm{num}(\ntot|\theta_\mathrm{tot}) = \frac{1}{\sqrt{2\pi}\sigma_\epsilon}\int_0^\infty \frac{(\theta_\mathrm{BG}+\epsilon\theta_\mathrm{S})^{\ntot}e^{-(\theta_\mathrm{BG}+\epsilon\theta_\mathrm{S})}}{\ntot!}\frac1\epsilon\exp\left[-\frac{1}{2}\left(\frac{\ln\epsilon}{\sigma_\epsilon}\right)^2\right]\mathrm{d}\epsilon \, ,
\end{equation}
where $\theta_{\mathrm{S}}$ is the predicted number of signal events, $\theta_\mathrm{BG}$ is the predicted number of background events, $\theta_\mathrm{tot} = \theta_{\mathrm{S}} + \theta_\mathrm{BG}$, $\epsilon$ is the rescaling variable assumed to have a log-normal distribution, and $\sigma_\epsilon$ is the fractional systematic error on the signal prediction (which sets the width of the distribution of $\epsilon$). The width $\sigma_\epsilon$ is the sum in quadrature of a theoretical error $\tau$ and the fractional uncertainty on the detector response. This treatment requires the selection of a single indicative systematic error on the effective volume, which is then applied identically at all muon energies.  When computing results, to be conservative we chose the largest systematic error on the effective volume over the entire range of detectable muon energies.  For the theoretical error $\tau$ we adopted a minimum of 5\% for WIMP masses $m_\chi \le 100$\,GeV to account for neglected higher order corrections and round-off errors, increasing to 50\% at $m_\chi=10$\,TeV as
\begin{equation}
\tau = 0.05\left(\frac{m_\chi}{100\,{\rm GeV}}\right)^{1/2}\,.
\end{equation}
This sliding scale is designed to encapsulate the increasing error with WIMP mass of predicted spectra from \textsf{DarkSUSY}, due to internal tables in which it interpolates results from \textsf{WIMPSim}.  Paper I and Refs.\ \cite{Conrad03,Scott09c} give further details and background on the number likelihood.

The expected distribution of incident neutrino energies ($\Etruei$) and angles ($\phitruei$) is given by $\diff^2 P / \diff \Etruei\,\diff \phitruei(\Etruei,\phitruei,\params)$, which is a prediction of the model parameters $\params$.
This separates into a weighted sum of the signal (S) and background (BG) contributions, so that Eq.\ \ref{unbinned_like} can be expressed as
\begin{equation}
\label{unbinned_like_simple}
\Like_\mathrm{unbin} \equiv \Like_\mathrm{num}(\ntot|\theta_\mathrm{tot})\prod_{i=1}^{\ntot} \left(f_\mathrm{S}\Like_{\mathrm{S},i}+f_\mathrm{BG}\Like_{\mathrm{BG},i}\right)\,,
\end{equation}
where $f_\mathrm{S} \equiv \theta_\mathrm{S} / \theta_\mathrm{tot}$ and $f_\mathrm{BG} \equiv \theta_\mathrm{BG} / \theta_\mathrm{tot}$ are the fractions of the total expected events from signal and background, respectively, and
\begin{equation}
\label{x_like}
\Like_{X,i}(\Eobsi, \phiobsi|\params) \equiv \int_0^\pi\int_0^\infty Q(\Eobsi, \phiobsi | \Etruei, \phitruei)\frac{\diff^2 P_X}{\diff \Etruei\,\diff \phitruei}(\Etruei,\phitruei,\params)\,\diff \Etruei\, \diff \phitruei\,,
\end{equation}
gives the signal ($X=\mathrm{S}$) and background likelihoods ($X=\mathrm{BG}$).

\subsection{Background likelihood}

The calculation of the background likelihood component follows the treatment in Paper I closely: the integral in Eq.\ \ref{x_like} for $X=\mathrm{BG}$ is the actual observed background, which is independent of the model parameters $\params$.  Within the zenith angle range considered in this analysis, to a very good approximation, the background spectrum and angular distributions are not correlated.  $\Like_{\mathrm{BG},i}$ can then be written as
\begin{equation}
\Like_{\mathrm{BG},i}(\Eobsi, \phiobsi) = \frac{\diff P_\mathrm{BG}}{\diff \Eobsi}(\Eobsi)\frac{\diff P_\mathrm{BG}}{\diff \phiobsi}(\phiobsi)\,,
\end{equation}
where $\diff P_\mathrm{BG}/\diff \Eobsi$ and $\diff P_\mathrm{BG}/\diff \phiobsi$ are the observed $N^{\rm c}_{\rm chan}$ and angular distributions of the background, respectively (Sec.\ \ref{icbackground}).  The expected number of background events $\theta_\mathrm{BG}$ used to calculate the background fraction $f_\mathrm{BG}$ refers to the events contained in the angular cut $\phi'_\mathrm{cut}$ around the solar position.

\subsection{Signal likelihood}

In order to take into account the distribution of production angles in calculating the signal likelihood, the integral in Eq.\ \ref{x_like} for $X=\mathrm{S}$ should be expressed in terms of the kinematics of the produced muons.  In Eq.\ \ref{x_like} this integrand is the product of the predicted arrival probability of a neutrino of a given energy and arrival angle, and the detector response to it.  We express this as the product of the predicted differential flux of incoming neutrinos ($\diff^2\Phi_\nu/\diff\Etruei\,\diff\phitruei$), the exposure time of the observation ($t_\mathrm{exp}$), the effective differential cross-section for neutrino conversion into muons in the ice ($\diff^2\Sigma_{\nu\to\mu}/\diff\Emui\,\diff\phimui$), and the response of the detector to muon-conversion events ($Q_\mu$).
We then integrate over the distribution of muon energies and angles that might be created in the interaction, so as to recover a pure function of the neutrino properties (as the theoretical predictions of different dark matter models $\params$ are given at neutrino level).
We divide by the expected number of signal events $\theta_\mathrm{S}$ inside the angular cut cone, in order to normalise the integral of the resulting probability distribution to unity.
We also multiply by a bias factor $f^{\rm b}(E)$, which is an analysis-dependent function of the neutrino energy.  $\theta_\mathrm{S}$ and $f^{\rm b}(E)$ are discussed in detail in Sec.\ \ref{eventrate}.
Finally, we add the contributions of both incoming neutrinos and antineutrinos, giving:
\begin{align}
\label{signal_like}
Q(\Eobsi, &\phiobsi | \Etruei, \phitruei) \frac{\diff^2 P_\mathrm{S}}{\diff \Etruei\,\diff \phitruei}(\Etruei,\phitruei,\params) = \nonumber\\
\frac{t_\mathrm{exp}}{\theta_\mathrm{S}}\Bigg[&\frac{\diff^2\Phi_\nu}{\diff\Etruei\,\diff\phitruei}(\Etruei,\phitruei,\params) f_\nu^{\rm b}(E) \int_0^\infty \int_\phitruei^{\phitruei+\pi} Q_\mu(\Eobsi,\phiobsi|\Emui,\phimui) \frac{\diff^2\Sigma_{\nu\to\mu}}{\diff\Emui\,\diff\phimui}(\Emui,\phimui|\Etruei,\phitruei)\,\diff \phimui\,\diff \Emui\nonumber\\
+&\frac{\diff^2\Phi_{\bar\nu}}{\diff\Etruei\,\diff\phitruei}(\Etruei,\phitruei,\params) f_{\bar\nu}^{\rm b}(E) \int_0^\infty \int_\phitruei^{\phitruei+\pi} Q_{\bar\mu}(\Eobsi,\phiobsi|\Emubari,\phimubari) \frac{\diff^2\Sigma_{\bar\nu\to\bar\mu}}{\diff\Emubari\,\diff\phimubari}(\Emubari,\phimubari|\Etruei,\phitruei)\,\diff \phimubari\,\diff \Emubari\Bigg]\,.
\end{align}
Here $\phimui$ is the angle of the produced muon relative to the Sun, $\Emui$ is its energy, and barred quantities are the equivalent measures for anti-particles.

The angular component of the signal prediction is a delta function at the solar position,
\begin{equation}
\frac{\diff^2\Phi_\nu}{\diff\Etruei\,\diff\phitruei}(\Etruei,\phitruei,\params) = \frac{\diff\Phi_\nu}{\diff\Etruei}(\Etruei,\params) \delta(\phitruei)\,,
\end{equation}
so the integral of Eq.\ \ref{signal_like} over $\phitruei$ (required by Eq.\ \ref{x_like} in order to obtain the signal likelihood) can be done analytically.  We then find
\begin{multline}
\label{signal_like_simpler}
\Like_{\mathrm{S},i}(\Eobsi, \phiobsi|\params) = \\
\frac{t_\mathrm{exp}}{\theta_\mathrm{S}}\sum_{\nu,\bar\nu}\int_0^\infty \frac{\diff\Phi_\nu}{\diff\Etruei}(\Etruei,\params) f_\nu^{\rm b}(E) \int_0^\infty \int_0^\pi Q_\mu(\Eobsi,\phiobsi|\Emui,\phimui) \frac{\diff^2\Sigma_{\nu\to\mu}}{\diff\Emui\,\diff\phimui}(\Emui,\phimui|\Etruei,0)\,\diff \phimui\,\diff \Emui\,\diff \Etruei\,,
\end{multline}
where the sum indicates that the corresponding antiparticle expression must also be included.  With $\phitruei=0$, the true muon arrival angle relative to the Sun, $\phimui$, becomes identical to the microscopic muon production angle in the frame where the target nucleus is at rest.  The value of this angle depends on the incoming neutrino energy and the outgoing muon energy, as well as the momentum carried by the parton within the nucleon with which the neutrino interacts.  It can be written as
\begin{equation}
\cos\phimui(x,y,\Etruei) = \left(\Emui^2 - m_\mu^2\right)^{-1/2}\left(\Emui - m_\mathrm{N}xy - m_\mu^2/2\Etruei \right)\,,
\end{equation}
where $m_\mathrm{N}$ refers to the mass of the nucleon involved. The Bj\"orken scaling variable $x$ indicates the fraction of the nucleonic momentum carried by the parton involved in the interaction.  By definition, $x$ varies between 0 and 1, as does the other Bj\"orken variable $y = 1-\Emui/\Etruei$.  Together, $x$ and $y$ provide a convenient and well-bounded way to express the dependence of the neutrino interaction cross-sections on the outgoing muon energy and angle.  We therefore trade $\Emui$ and $\phimui$ for $x$ and $y$, so that $\Like_{\mathrm{S},i}(\Eobsi, \phiobsi|\params)$ becomes
\begin{equation}
\label{signal_like_simple}
\frac{t_\mathrm{exp}}{\theta_\mathrm{S}}\sum_{\nu,\bar\nu}\int_0^\infty \frac{\diff\Phi_\nu}{\diff\Etruei}(\Etruei,\params) f_\nu^{\rm b}(E) \iint_0^1 Q_\mu\left(\Eobsi,\phiobsi|\Emui,\phimui\right) \frac{\diff^2\Sigma_{\nu\to\mu}}{\diff x\,\diff y}(x,y|\Etruei)\,\diff x\,\diff y\,\diff\Etruei\,,
\end{equation}
remembering that $\Emui = \Emui(y,\Etruei)$ and $\phimui = \phimui(x,y,\Etruei)$.  For each observed event inside our analysis cone, we precompute the inner double integral of Eq.\ \ref{signal_like_simple} for a set of 50 logarithmically-spaced neutrino energies per decade over the range $0.5 \le \log_{10}(E/\mathrm{GeV}) \le 4.0$.  To obtain the contribution of the predicted signal to the total likelihood for that event, we re-weight these `partial likelihoods' according to the predicted neutrino spectrum for each model $\params$, as well as the bias factor $f^{\rm b}$.  To allow for a fast and straightforward application to any theoretical neutrino spectrum, we provide the partial likelihoods for the 79-string IceCube analysis in \textsf{nulike}, precomputed, along with routines for computing the bias factors $f^{\rm b}$.  We also provide the underlying event data online \cite{filesloc} and a utility within \textsf{nulike} that can precompute and save the partial likelihoods from any other neutrino telescope, provided the data are in the same format.

The effective differential conversion cross-section is given by
\begin{equation}
\label{diffxsec}
\frac{\diff^2\Sigma_{\nu\to\mu}}{\diff x\,\diff y}(x,y|\Etruei) =  V_{\rm eff}(\Emui) \sum_{N=p,n} n_N \frac{\diff^2\sigma_{\nu\to\mu,N}}{\diff x\,\diff y}(x,y | \Etruei)\,,
\end{equation}
where it should again be understood that $\Emui=\Emui(y,\Etruei)$.  The replacement $\{\nu,\mu\}\to\{\bar\nu,\bar\mu\}$ provides the corresponding expression for $\diff^2\Sigma_{\bar\nu\to\bar\mu}/\diff x\,\diff y$.  This is the product of the number density $n_N$ of nucleon species $N$ (proton or neutron) in the detector, the effective volume $V_\mathrm{eff}(\Emui)$ of the detector for muon or anti-muon conversion events, and $\diff^2\sigma_{\nu\to\mu,N}/\diff x\,\diff y$, the microscopic differential cross-section for muon production by charged-current interactions.  $V_\mathrm{eff}$ is the same for both muons and anti-muons.  In contrast, the differential cross-sections differ for particles and antiparticles.  These are known from the theory of weak interactions, up to a dependence on the parton distributions for $x$.  We obtain these from \textsf{nusigma} \cite{nusigma}, which by default relies on the CTEQ6-DIS parton distribution functions \cite{CTEQ6}.  Users of \textsf{nulike} who prefer other parton distributions can simply switch those employed by \textsf{nusigma} and recompute the partial likelihoods.

\subsection{Detector response}

Based on the observation that the angular and spectral ($N^{\rm c}_{\rm chan}$) distributions of detected events are essentially uncorrelated across the sky (Sec.\ \ref{icbackground}), we assume that the detector response to events producing muons factorises into the product
\begin{equation}
\label{Qmu}
Q_\mu(\Eobsi,\phiobsi|\Emui,\phimui) = E_\mathrm{disp}(\Eobsi|\Emui){\rm PSF}(\phiobsi|\phimui,\Emui)\,.
\end{equation}
Here $E_\mathrm{disp}(\Eobsi|\Emui)$ is the energy dispersion of the detector and PSF$(\phiobsi|\phimui,\Emui)$ is its point spread function, assuming these to be identical for muons and antimuons.  The energy dispersion is the $N^{\rm c}_{\rm chan}$ response to events that produce muons of a given energy (in contrast to the neutrino $N_{\rm chan}$ response that we employed in Paper I).  We obtained this from IceCube detector Monte Carlo simulations (Sec.\ \ref{icenergy}).

The uncertainty in the muon reconstruction direction is given on a per-event basis by the IceCube paraboloid sigma $\sigma_{{\rm para},i}$ for the $i$th event (Sec.\ \ref{icangular}), which accounts for the dependence of the PSF on the incoming muon energy.  To obtain the PSF in terms of $\phiobsi$ and $\phi_\mu$, we shift from the coordinate system centred on the true muon arrival direction (i.e. $\Delta=0$ in Eq.~\ref{PSF}) to the one with the Sun at the origin ($\phi'=0$), integrating over all azimuths to obtain
\begin{equation}
\label{offcentrepsf}
{\rm PSF}(\phiobsi|\phi_\mu,\Emui) = \frac{\phiobsi}{\sigma_{{\rm para},i}^{\phantom{{\rm para},i}2}} \exp\left[-\frac{{\phiobsi}^2+{\phi_\mu}^2}{2\sigma_{{\rm para},i}^{\phantom{{\rm para},i}2}}\right]I_0\left(\frac{\phi_\mu\phiobsi}{\sigma_{{\rm para},i}^{\phantom{{\rm para},i}2}}\right)\,,
\end{equation}
where $I_0$ is the lowest-order modified Bessel function of the first kind.\footnote{The definition of the PSF here differs from Paper\ I, as in the current paper we allow for differences between the neutrino and muon angles, perform the co-ordinate shift and normalise over the whole sky.  Previously, we normalised over the analysis cut cone instead of the full sky, and normalised with respect to the allowed ranges of the true direction and its deviation from the reconstructed one, rather than by simply considering the permitted values of the reconstructed angle like we do here.  This marginally degraded the sensitivity of the previous analysis, although the effect was negligible in comparison to the experimental uncertainty.}

\subsection{Predicted event rate}
\label{eventrate}

The total predicted number of signal events $\theta_\mathrm{S}$ follows similarly to Eq.\ \ref{signal_like_simple} as the sum of the predicted number of neutrino-initiated signal events
\begin{equation}
\label{signal}
\theta_{\mathrm{S},\nu}(\params) = t_\mathrm{exp}\int_0^\infty \frac{\diff\Phi_\nu}{\diff\Etruei}(\Etruei,\params) f_\nu^{\rm b}(E) \iint_0^1 L(\phimui,\Emui,\phi'_\mathrm{cut})\frac{\diff^2\Sigma_{\nu\to\mu}}{\diff x\,\diff y}(x,y|\Etruei)\,\diff x\,\diff y\,\diff \Etruei\,,
\end{equation}
and the corresponding quantity $\theta_{\mathrm{S},\bar\nu}$ for anti-neutrinos.  Again, we remind the reader that $\phimui$ and $\Emui$ are functions of $x$, $y$ and $E$.  The only difference here with respect to what one would naively read off Eq.\ \ref{signal_like_simple} is the factor $L(\phimui,\Emui,\phi'_\mathrm{cut})$, a dimensionless, energy-dependent angular loss factor that is independent of the muon charge.  $L$ corrects for neutrinos that originate from the direction of the Sun but produce muons that are ultimately reconstructed as arriving from outside the analysis cut cone ($\phiobsi>\phi'_\mathrm{cut}$).  Similarly to Paper I, we use the mean angular error of IceCube ($\sigma_\mu$; cf.\ Sec.\ \ref{icangular}) to calculate $L$, integrating the PSF over the analysis cut cone to give
\begin{equation}
\label{anglossfactor}
L(\phimui, \Emui,\phi'_\mathrm{cut}) = \int_0^{\phi'_\mathrm{cut}} \frac{\phi'}{\sigma_\mu^2} \exp\left[-\frac{{\phi'}^2+{\phi_\mu}^2}{2\sigma_\mu^{2}}\right]I_0\left(\frac{\phi_\mu\phi'}{\sigma_\mu^{2}}\right) \diff \phi' \equiv P_1\left(\frac{\phi_\mu}{\sigma_\mu},\frac{\phi'_\mathrm{cut}}{\sigma_\mu}\right)\,.
\end{equation}
This is known as the Marcum $P$-function or Complementary Marcum $Q$-function, which we evaluate with the code of Ref.\ \cite{Gil14}.

There are two crucial differences here as compared to Paper I.  The first is that $L$ is a muon-level correction factor, expressed in terms of the muon energy and the width of the muon-level angular uncertainty $\sigma_\mu(\Emui)$, not the corresponding neutrino quantities.  The other is that because of the non-zero muon production angle, the off-centre PSF (Eq.\ \ref{offcentrepsf}) must be used instead of the central distribution (Eq.\ \ref{PSF}).

The mean angular error is the correct PSF width to use in Eq.\ \ref{anglossfactor}, because we are interested in determining \textit{a priori} what fraction of incoming neutrinos with a given energy should be absent from the final set of observed events, due to the chosen angular cut.  This is in contrast to the case of the contribution to the partial likelihood coming from the detector response (Eq.\ \ref{offcentrepsf}), where we are interested in the probability that \textit{a given event} originated from the Sun, where the event-level paraboloid $\sigma_{{\rm para},i}$ should be preferred.

\subsection{Bias factor calculation}
\label{biasFactor}

The inner double integral in Eq.\ \ref{signal} gives the \textit{unbiased} neutrino effective area for this analysis.  It differs from the effective area derived in the standard 79-string analysis \cite{IC79} in two important ways.  First, it includes the factor $L$, to account for the angular loss due to our analysis cut cone around the solar position.  Second, it implicitly assumes that all muons of a given energy are equally likely to pass the original analysis cuts used in the 79-string analysis.  In reality, low-energy muons created by high-energy neutrinos are, for example, far more likely to appear in the final event sample than muons of the same energy created by low-energy neutrinos.  This is due to the additional light deposited in the detector from the hadronic recoil in the case of a higher-energy neutrino, and the analysis cuts placed on quantities such as the absolute number of activated DOMs.

This departure from a perfect mapping between the properties of a muon and its probability of ending up in the final event sample constitutes a bias that depends on the neutrino energy.  This is precisely the reason for the bias factor $f^{\rm b}(E)$ in the preceding expressions, which accounts for the departure of the the event sample from the minimum bias expectation.  To quantify this effect, we take the ratio
\begin{equation}
f^{\rm b}(E) =  A_{\rm eff}(E) \left[ \iint_0^1 \frac{\diff^2\Sigma_{\nu\to\mu}}{\diff x\,\diff y}(x,y|\Etruei)\,\diff x\,\diff y, \right]^{-1}
\end{equation}
of the original 79-string effective area $A_{\rm eff}(E)$ to the unbiased effective area calculated without the angular correction $L$.  The final effective area in this paper is the product of the bias factor $f^{\rm b}$ and the unbiased effective area \textit{with} the angular correction $L$.  In this way, our analysis is fully consistent with the original 79-string effective area by construction, and accounts for both the bias and the angular cut cone at the same time.

\begin{figure}[tp]
\centering
\includegraphics{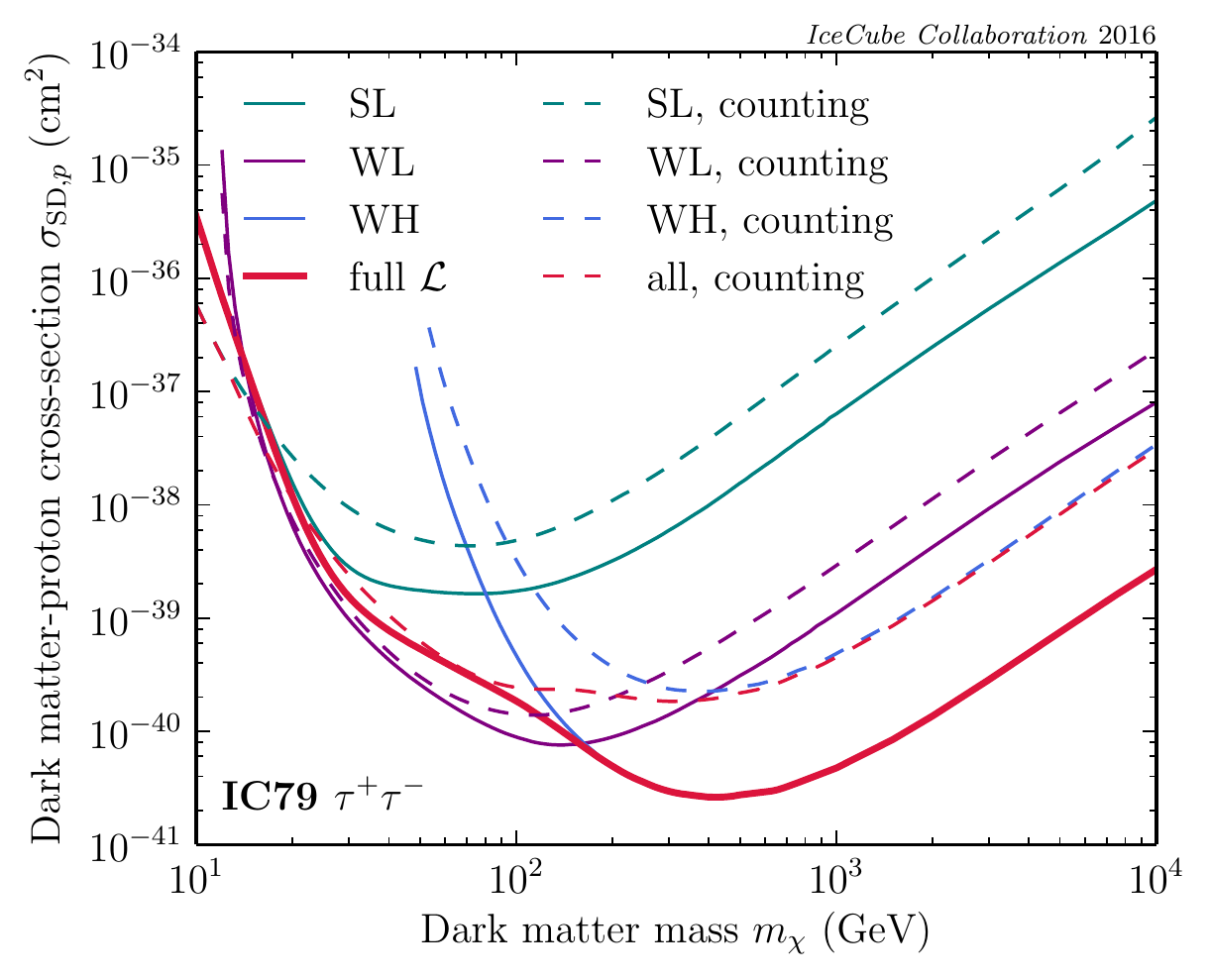}
\caption{Limits on dark matter annihilation in the Sun using an analysis that takes into account neutrino energy information.  We show limits separately for the three different IC79 event samples SL (summer low) WL (winter low) and WH (winter high) and their combination.  The difference between dashed and solid lines indicates the improvement gained by moving from a simple counts-based number likelihood to a full unbinned one, incorporating the number of events, their arrival directions and energies.  The full limit is weaker than the WL sample taken alone at low masses, because the SL sample exhibits a weak excess ($<$$2\sigma$ local significance) of events above background expectation, not borne out in the WL sample. Here we have assumed an annihilation cross-section of \sv\ entirely into $\tau^+\tau^-$ final states.}
\label{fig::analyses}
\end{figure}

To facilitate the use of other neutrino spectra, we provide unbiased effective areas precomputed in \textsf{nulike} for the three 79-string IceCube event selections, both with and without the angular correction $L$.  We also provide the routines necessary to repeat the computations for any other dataset.  In final likelihood mode, the user can choose to have \textsf{nulike} work with user-supplied bias factors, or use the unbiased effective areas to automatically determine the bias factors.

\begin{figure}[tp]
\centering
\includegraphics{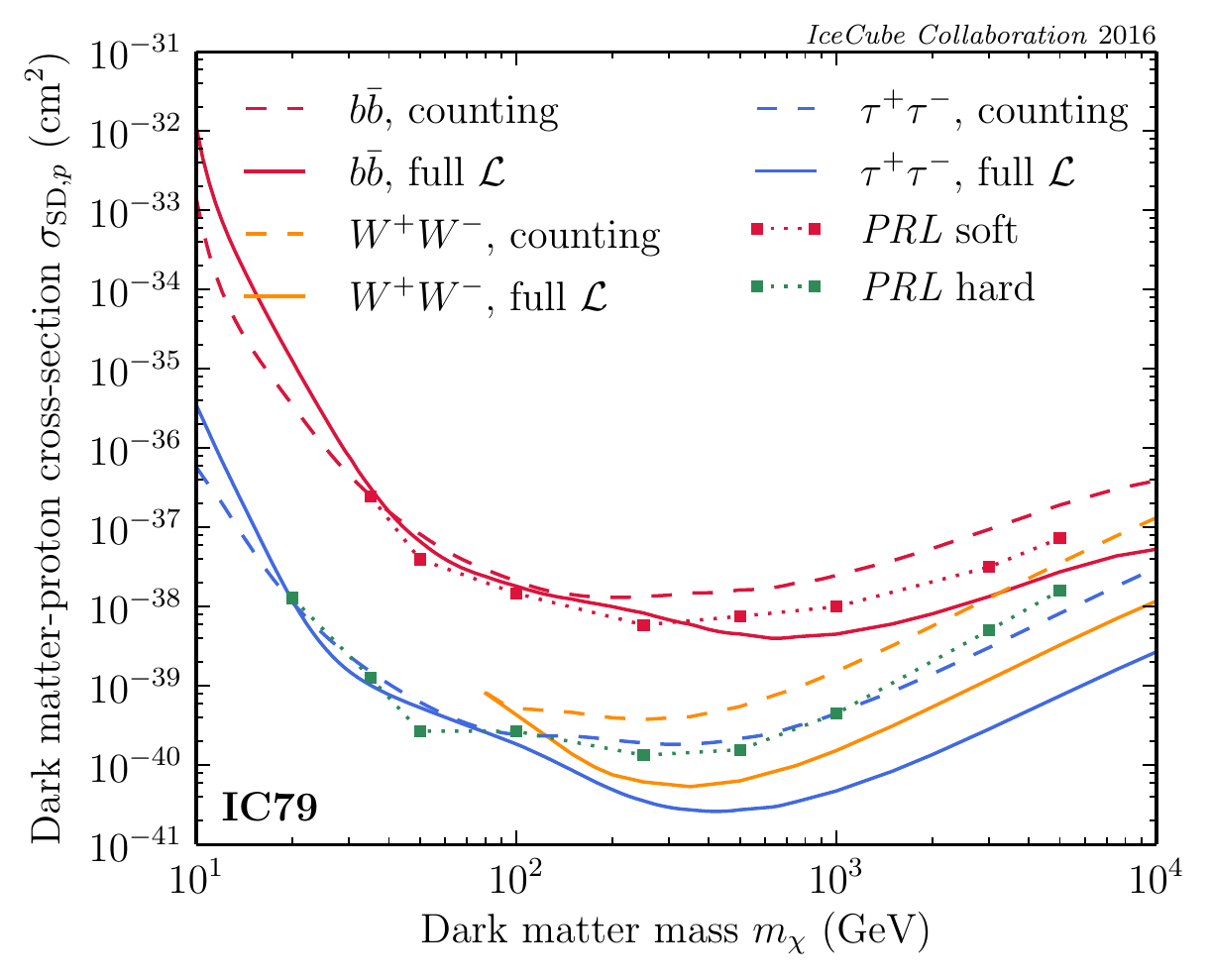}
\caption{Limits on the spin-dependent WIMP-proton cross-section from IC79 using the improved likelihood, for the canonical soft ($b\bar{b}$) and hard ($W^+W^-$ and $\tau^+\tau^-$) annihilation channels often seen in SUSY models.  Here we compare to the limits from the original IC79 analysis (`PRL'; \cite{IC79}); note that the previous `hard' channel limit is $W^+W^-$ above the $W$ mass, but $\tau^+\tau^-$ below it.  The addition of energy information provides an improvement of up to a factor of 4 at high WIMP masses over the previous analysis, whereas the limits are in excellent agreement for low WIMP masses.  Here we have assumed an annihilation cross-section of \sv.}
\label{fig::newlimits}
\end{figure}

\begin{figure}[tp]
\centering
\includegraphics{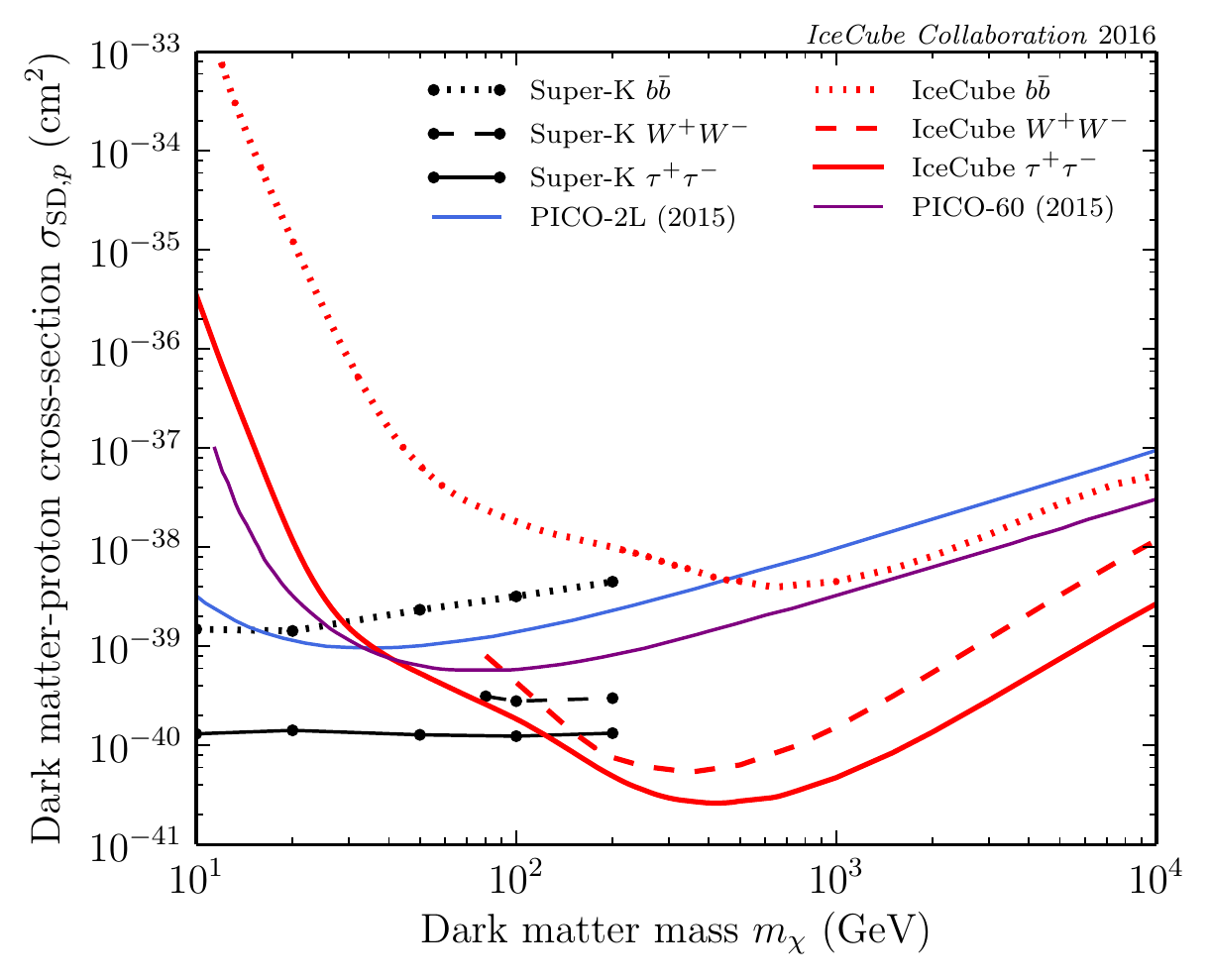}
\caption{Comparison of our limits with the latest constraints from Super-Kamiokande \cite{SuperK15} and PICO \cite{PICO2L,PICO60}.  Depending on the annihilation channel, IceCube provides the strongest limits above WIMP masses of $\sim$100--200\,GeV.  Super-K is more sensitive at the lowest masses.  If the annihilation spectrum is soft or heavily suppressed, the PICO experiment provides stronger limits than neutrino telescopes; other direct limits are weaker. Here we have assumed an annihilation cross-section of \sv\ for deriving IceCube limits; Super-K limits assume complete equilibrium between capture and annihilation in the Sun.}
\label{fig::superk}
\end{figure}

\begin{figure}[tp]
\centering
\includegraphics{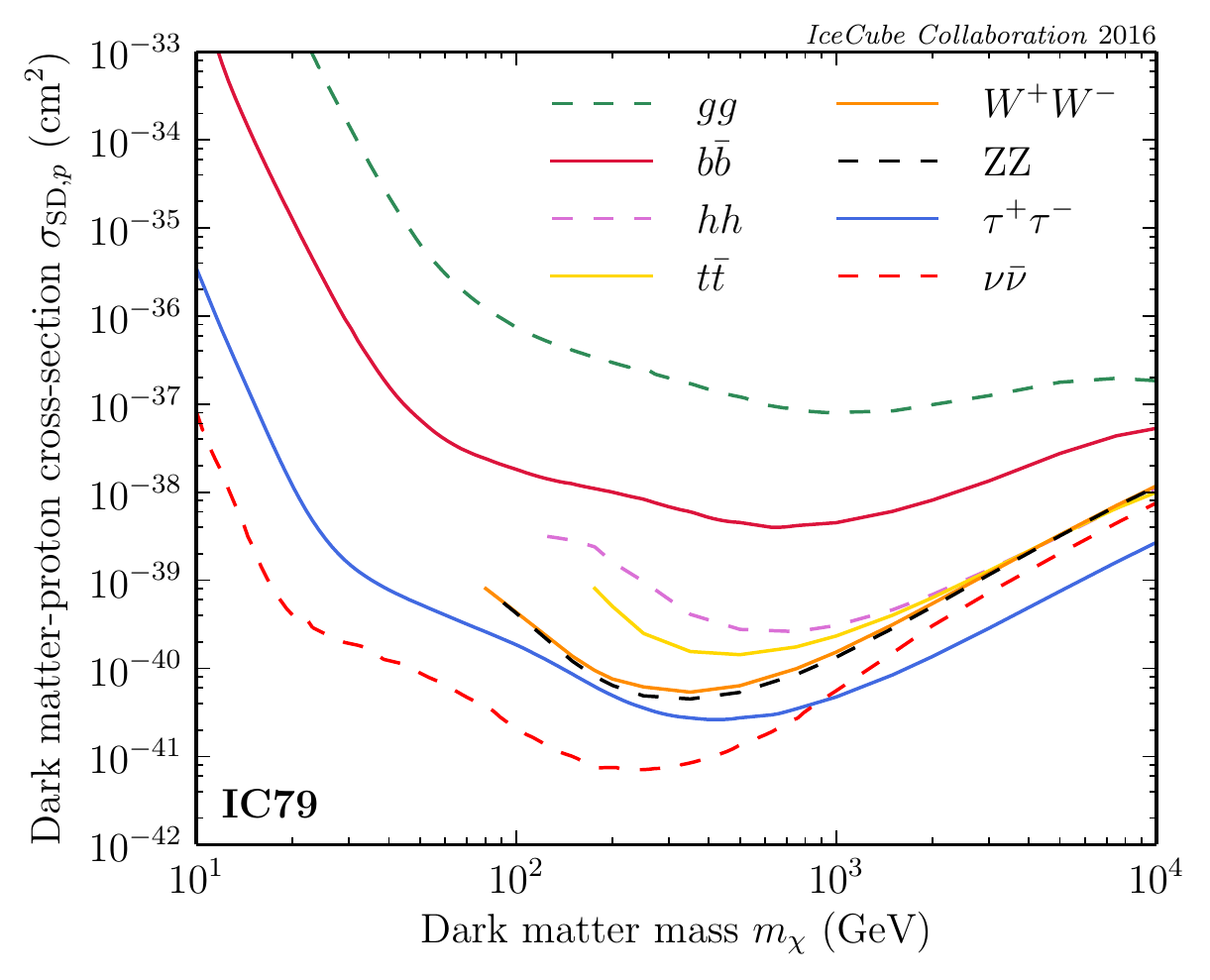}
\caption{Limits on the spin-dependent WIMP-proton cross-section from IC79, for a range of different annihilation final states.  The canonical hard ($W^+W^-$ and $\tau^+\tau^-$) and soft ($b\bar{b}$) channels bracket the possible limits for different models reasonably well. More extreme channels (hardest: $\nu\bar{\nu}$, softest: $gg$) less often found in SUSY can lead to even stronger or weaker constraints.  For the $\nu\bar{\nu}$ channel we have assumed equal branching fractions for all three neutrino flavours. The ability to easily and quickly compute full limits for any combination of final states is a particular feature of the method and tools we present in this paper.  As a convenience, datafiles for all curves in this figure are available precomputed in the \textsf{nulike} download\protect\footnote{\href{http://nulike.hepforge.org}{http://nulike.hepforge.org}}.}
\label{fig::all_channels}
\end{figure}

\section{Improved limits on WIMP dark matter}
\label{results_wimp}

Figures\ \ref{fig::analyses}--\ref{fig::all_channels} show the 90\% confidence level (CL) limits on simple effective WIMP DM models computed using IC79 data (Sec.\ \ref{data}) and the \textsf{nulike 1.0.0} implementation of the likelihood described in Sec.\ \ref{likelihood}.  We use the $\Delta\ln\Like$ relative to the background-only prediction as the test statistic, summed over the three event selections, conditioning on all parameters except the cross-section to leave only a single degree of freedom.  The distribution of this test statistic is very close to $\chi^2$, as shown in previous analyses by explicit Monte Carlo \cite{IC79, DanningerThesis}; this allows CLs to be determined by standard $\Delta\chi^2$ methods.

For all limits in this section, we assume that DM annihilates exclusively to some specific final state, with a canonical thermal annihilation cross-section \sv.  For all but the highest WIMP masses and lowest scattering cross-sections, these models have reached equilibrium between capture and annihilation in the Sun.  We do not assume equilibrium in our calculations however, as is often done.  We use \textsf{DarkSUSY 5.1.3} to compute the predicted neutrino spectrum at the detector for each model, and to solve for the present-day DM population in the Sun.  We adopt the standard halo model and default nuclear matrix elements as implemented in \textsf{DarkSUSY}; see discussions in Refs.\ \cite{Akrami11DD, Cline13b}.

Fig.\ \ref{fig::analyses} presents the limits on the spin-dependent WIMP-proton cross-section imposed by the three different IC79 event samples: WH, WL and SL individually, and in combination.  As an example, here we show limits corresponding to annihilation solely to $\tau^+\tau^-$ final states.  As expected \cite{IC79, DanningerThesis}, the SL and WL samples dominate the sensitivity at low WIMP masses.  For comparison, we also show limits based on the number likelihood (Eq.\ \ref{number}) alone, neglecting all event-level information.  For the cut cone that we use (40 degrees for WL and SL, 20 degrees for WH), considering the arrival directions and energies of neutrino events provides up to a factor of 20 improvement in the resulting limits.

At high masses, the combined limit in Fig.\ \ref{fig::analyses} essentially tracks the exclusion curve of the WH sample, which is orders of magnitude more sensitive than the WL and SL samples in this region of parameter space.  At masses below 100\,GeV however, where SL and WL both play significant roles, the combined limit is slightly weaker than the limit obtained by considering the WL sample alone.  This is because the SL sample exhibits a weak excess above the background expectation inside the analysis cut cone that is not replicated in the WL sample: 819 observed events as compared to 770 predicted in the analysis cone from background alone.

In Fig.\ \ref{fig::newlimits} we compare these new limits to the previous 79-string IceCube constraints on hard and soft annihilation channels.  To allow a reasonable comparison, here we show limits for $b\bar{b}$, $W^+W^-$ and $\tau^+\tau^-$ final states, matching what was used in the previous analysis (`soft channel' = $b\bar{b}$, `hard channel' = $W^+W^-$ for $m_\chi > m_W$ and $\tau^+\tau^-$ for $m_\chi < m_W$).  The previous analysis used the same data as we use here, except that it did not include event energy information in the likelihood function.  At low masses, the analysis agrees with the previous one, indicating that the energy information adds little information.  Including the event-level energy information has the most impact at high WIMP mass, making use of the relatively good energy resolution of IceCube at high muon energies.  The limits in Fig.\ \ref{fig::newlimits} are up to a factor of 4 stronger than the previous analysis at multi-TeV masses. The latest update of \textsf{WIMPSim} fixes an issue with propagation of neutrinos in the Sun that affected the version used to derive the original IC79 limits~\cite{IC79}. This resulted in conservative limits for WIMP masses above $\sim$500\,GeV, ranging from a factor of 1.05 at 500\,GeV to 1.2 at 1\,TeV and up to 1.5 at 5\,TeV for the $W^+W^-$ and $\tau^+\tau^-$ final states. Improvements beyond those factors are due to the improved analysis method in this paper.

Fig.\ \ref{fig::superk} compares these limits to other searches for spin-dependent DM-proton scattering, both from the Sun and direct detection experiments.  The 79-string IceCube data provide the strongest limits of any search for all masses above $\sim$100--200\,GeV (the exact value depends on the annihilation channel).  Super-Kamiokande \cite{SuperK15} is the most sensitive experiment at all lower masses. Limits from direct detection \cite{PICO2L,PICO60} are weaker, except in the case of DM with soft or suppressed annihilation spectra, in which case the PICO experiment \cite{PICO2L,PICO60} is the most constraining.  Indirect DM searches by Antares~\cite{Antares_2016} and Baksan~\cite{Baksan13} have set less stringent limits on the spin-dependent DM-proton scattering and are consequently not included in Fig.\ \ref{fig::superk}.

Figure\ \ref{fig::all_channels} shows new limits for all major two-body annihilation final states.  Annihilation to either electroweak gauge boson final state is more or less equivalent, as $W$ and $Z$ have around the same mass and couplings to the rest of the SM, and consequently yield very similar neutrino spectra.  We don't show $hZ$, but we have checked that it indeed lies mid-way between $hh$ and $ZZ$, as expected.

As expected, most channels are indeed bracketed by the canonical `hard' and `soft' channels. The exceptions to this are gluon final states, where spectra are especially soft and limits particularly weak, and neutrino final states, which give very strong limits because they are monochromatic at the source.  In the neutrino case, the monochromatic source spectrum means that of all final states, annihilation to neutrinos tracks the actual neutrino effective area most closely, with the only deviation from a monochromatic spectrum at the detector coming from reprocessing in the Sun following prompt production at the DM mass.  This is also why the neutrino-channel limits at masses above one TeV become weaker than those from the $\tau^+\tau^-$ channel: as a channel with an extremely hard annihilation spectrum, most of the neutrinos produced are close to the DM mass, and are therefore absorbed in the Sun.  This is a general feature of all channels above one TeV: soft and hard channels begin to swap character in terms of the limits, as softer channels actually produce more neutrinos able to make it out of the Sun and to the detector.  This effect can also be seen in the gluon channel limits, which become \textit{stronger} as the mass increases past $\sim$7\,TeV, as enough of the resulting very low-energy neutrinos are pulled into the observable energy window from below to counteract the slight increase in the number of neutrinos above one TeV that never make it out of the Sun.

\begin{figure}[tp]
\centering
\includegraphics{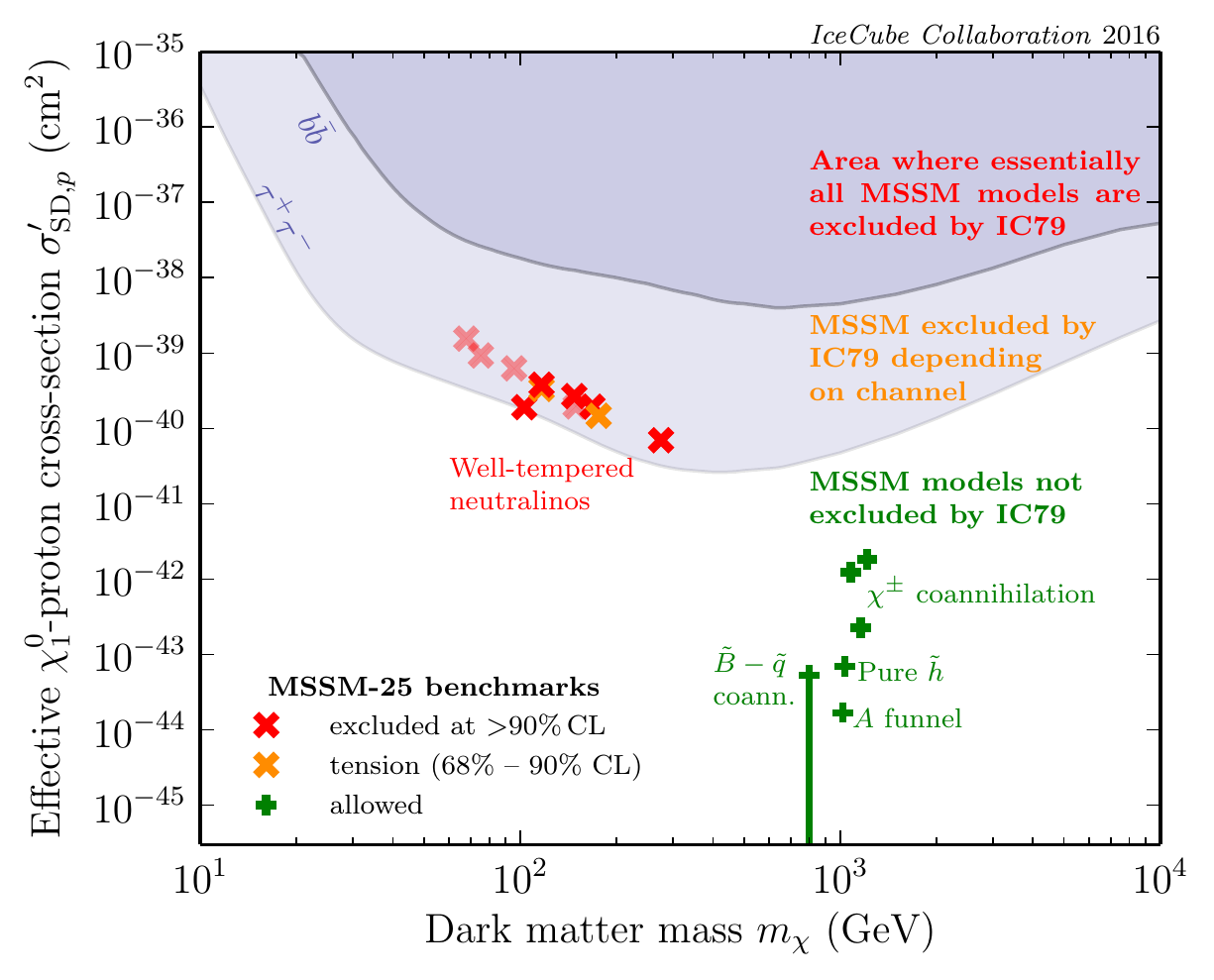}
\caption{Implications of the new IC79 analysis for benchmark models in the MSSM-25.  Models shown with solid red crosses are excluded for the first time by IC79.  Faded red symbols are excluded by both IC79 and recent LUX spin-dependent bounds \cite{LUXCalc, LUX13}.  These all correspond to so-called `well-tempered' neutralinos, which exhibit a mixed gaugino-Higgsino character.  Solid orange crosses indicate models in tension with IC79 data at more than 1$\sigma$ (but excluded at less than 90\% CL).  Green plus symbols indicate models not constrained by IC79, labelled according to the dominant characteristic determining their relic density.  The vertical green line corresponds to a benchmark `spoke' of models \cite{SM13}, where the correct relic density is obtained by bino-squark co-annihilation.  Benchmarks are from the MSSM-25 and MSSM-19 (`pMSSM'; a subset of the MSSM-25) scans of Refs.\ \cite{Silverwood12, SM13}, and correspond to models allowed by LHC, relic density and other direct and indirect constraints. Benchmark scattering cross-sections are rescaled for the neutralino relic density, and the shaded regions are indicative only; these assume pure spin-dependent scattering and annihilation to the canonical `hard' and `soft' channels often seen in the MSSM (even though harder and softer spectra are also possible within the MSSM).}
\label{fig::susy}
\end{figure}

\section{Implications for MSSM benchmarks}
\label{results_mssm}

In this section we use the new IceCube 79-string likelihood to test a number of models of weak-scale supersymmetry, employing the same test statistic as in Sec.\ \ref{results_wimp}.  Here we focus on the MSSM-25, a 25-parameter, weak-scale parameterisation of the minimal supersymmetric standard model (MSSM; see Ref.\ \cite{Silverwood12} for details).  This contains the MSSM-19, otherwise known to as the `phenomenological' (p)MSSM, as a subspace.

Fig.\ \ref{fig::susy} shows some MSSM-25 benchmark models from the study of Ref.\ \cite{Silverwood12}, selected by requiring models with large spin-dependent scattering cross-sections. To give a broader indication of the possibilities in the MSSM, Fig.\ \ref{fig::susy} also shows all models from the MSSM-19 benchmarking exercise of the Snowmass 2013 review \cite{SM13}, except for the Bino-stop co-annihilation benchmark, which is very similar to the Bino-squark benchmark in this plane.\footnote{We have reduced the Bino mass parameter $M_1$ in this benchmark from 868\,GeV to 800\,GeV, in order to make the neutralino the lightest SUSY particle when carrying out the calculations with \textsf{DarkSUSY}. This reduces the nuclear scattering cross-sections compared to Ref.\ \cite{SM13}, but the model is unconstrained by IceCube either way.}  Except for the models that we show with faded symbols (which we return to later) these models are all consistent with constraints from the LHC, flavour physics and the relic density of dark matter, as well as direct and indirect searches for dark matter.  The Snowmass 2013 benchmarks include a `spoke' of models extending along a single direction in parameter space from one specific benchmark, shown as a vertical line in Fig.\ \ref{fig::susy}.  We also show shaded bands between the strongest ($\tau^+\tau^-$) and weakest ($b\bar{b}$) limits for channels typically seen in the MSSM.  This gives some idea of where essentially all MSSM models are excluded regardless of annihilation channel (above the $b\bar{b}$ limit), and where only some models are excluded (between $b\bar{b}$ and $\tau^+\tau^-$), depending on their specific annihilation branching fractions to different final states.

We have colour-coded the individual models in Fig.\ \ref{fig::susy} by the extent to which they are excluded by the new IceCube limits, taking into account both spin-dependent and spin-independent scattering. We have also labelled different benchmark groups according to the means by which the neutralino achieves the appropriate relic density in the early Universe.  Many neutralino models are excluded for the first time by the new limits we present here (bright red crosses). Other models exhibit a tension with data at the 68--90\% confidence level (orange crosses).  These are `well-tempered' neutralino models, which exhibit a roughly even mixture of gaugino and Higgsino weak eigenstates, boosting their spin-dependent scattering cross-section without contributing too strongly to the spin-independent one.  Other benchmarks (green plus symbols), where the relic density is achieved by squark or chargino co-annihilation with the neutralino, resonant annihilation via the CP-odd Higgs, or by virtue of the large annihilation cross-section exhibited by pure Higgsinos, remain unconstrained by spin-dependent searches of any kind.

\begin{sidewaystable}[p]
\caption{Properties of the benchmark models shown in Fig.\ \protect\ref{fig::susy}.}
\label{bftable}
\scriptsize
\begin{tabular}{r@{\hspace{2mm}}c@{\hspace{2mm}}c@{\hspace{2mm}}c@{\hspace{2mm}}c@{\hspace{2mm}}c@{\hspace{2mm}}c@{\hspace{2mm}}c@{\hspace{2mm}}c@{\hspace{4mm}}l@{}r}
\hline
\multicolumn{1}{c}{$m_\chi$}  & $\langle\sigma v\rangle$ & $\Omega h^2$  & $\sigma_{\mathrm{SI}p}$   & $\sigma_{\mathrm{SD}p}$   & $\sigma^{'}_{\mathrm{SI}p}$   & $\sigma^{'}_{\mathrm{SD}p}$  & $C$ & $A$ & \multicolumn{1}{c}{Dominant annihilation final states} & \multicolumn{1}{c}{Excl.}\\
\multicolumn{1}{c}{(GeV)}  & (cm$^{-3}$\,s$^{-1}$) &  & (cm$^2$)   & (cm$^2$)   & (cm$^2$)   &  (cm$^2$)  & (s$^{-1}$)    & (s$^{-1}$)    & \multicolumn{1}{c}{(contributions $<$1\% not shown)} & \multicolumn{1}{c}{(\%$CL$)} \\
\hline
67.8   & 4.2 $\times10^{-28}$  & 0.123  & 4.1 $\times10^{-47}$  & 1.6 $\times10^{-39}$  & 4.1 $\times10^{-47}$  & 1.6 $\times10^{-39}$  & 7.9 $\times10^{23}$  & 4.0 $\times10^{23}$   & $\tau^+\tau^-$ (59\%), $b\bar{b}$ (16\%), $c\bar{c}$ (14\%), $gg$ (10\%), $Z\gamma$ (1\%)                                                                     & 100.0\\
75.3   & 6.4 $\times10^{-28}$  & 0.109  & 5.7 $\times10^{-47}$  & 1.0 $\times10^{-39}$  & 5.1 $\times10^{-47}$  & 9.3 $\times10^{-40}$  & 3.9 $\times10^{23}$  & 2.0 $\times10^{23}$   & $\tau^+\tau^-$ (71\%), $b\bar{b}$ (14\%), $c\bar{c}$ (6\%), $gg$ (5\%), $Z\gamma$ (3\%), $\gamma\gamma$ (1\%)                                                 & 100.0\\
95.6   & 1.6 $\times10^{-26}$  & 0.103  & 4.7 $\times10^{-48}$  & 7.4 $\times10^{-40}$  & 4.0 $\times10^{-48}$  & 6.3 $\times10^{-40}$  & 1.7 $\times10^{23}$  & 8.4 $\times10^{22}$   & $W^+W^-$ (85\%), $\tau^+\tau^-$ (12\%), $ZZ$ (3\%)                                                                                                             & 99.4 \\
102.9  & 1.9 $\times10^{-26}$  & 0.115  & 5.8 $\times10^{-46}$  & 2.0 $\times10^{-40}$  & 5.5 $\times10^{-46}$  & 1.9 $\times10^{-40}$  & 4.5 $\times10^{22}$  & 2.3 $\times10^{22}$   & $\tau^+\tau^-$ (93\%), $W^+W^-$ (5\%), $ZZ$ (2\%)                                                                                                             & 92.1 \\
116.6  & 2.0 $\times10^{-26}$  & 0.101  & 6.7 $\times10^{-46}$  & 4.0 $\times10^{-40}$  & 5.6 $\times10^{-46}$  & 3.3 $\times10^{-40}$  & 6.2 $\times10^{22}$  & 3.1 $\times10^{22}$   & $W^+W^-$ (53\%), $ZZ$ (30\%), $b\bar{b}$ (14\%), $Zh$ (2\%)                                                                                                     & 76.6 \\
116.6  & 1.7 $\times10^{-26}$  & 0.114  & 2.4 $\times10^{-46}$  & 4.0 $\times10^{-40}$  & 2.2 $\times10^{-46}$  & 3.8 $\times10^{-40}$  & 7.1 $\times10^{22}$  & 3.5 $\times10^{22}$   & $W^+W^-$ (62\%), $ZZ$ (35\%), $Zh$ (2\%), $b\bar{b}$ (1\%)                                                                                                    & 94.0 \\
147.5  & 1.8 $\times10^{-26}$  & 0.123  & 4.9 $\times10^{-48}$  & 2.7 $\times10^{-40}$  & 4.9 $\times10^{-48}$  & 2.7 $\times10^{-40}$  & 3.2 $\times10^{22}$  & 1.6 $\times10^{22}$   & $W^+W^-$ (52\%), $ZZ$ (36\%), $Zh$ (7\%), $b\bar{b}$ (4\%), $\tau^+\tau^-$ (1\%)                                                                              & 99.5 \\
149.0  & 2.6 $\times10^{-26}$  & 0.086  & 2.1 $\times10^{-44}$  & 2.8 $\times10^{-40}$  & 1.5 $\times10^{-44}$  & 2.0 $\times10^{-40}$  & 2.4 $\times10^{22}$  & 1.2 $\times10^{22}$   & $W^+W^-$ (53\%), $ZZ$ (37\%), $Zh$ (6\%), $b\bar{b}$ (2\%), $\tau^+\tau^-$ (1\%)                                                                              & 98.1 \\
168.1  & 1.7 $\times10^{-26}$  & 0.121  & 1.3 $\times10^{-45}$  & 1.9 $\times10^{-40}$  & 1.3 $\times10^{-45}$  & 1.9 $\times10^{-40}$  & 1.8 $\times10^{22}$  & 9.0 $\times10^{21}$   & $W^+W^-$ (51\%), $ZZ$ (39\%), $Zh$ (5\%), $b\bar{b}$ (3\%), $\tau^+\tau^-$ (2\%)                                                                              & 99.7 \\
175.9  & 1.7 $\times10^{-26}$  & 0.121  & 7.8 $\times10^{-48}$  & 1.5 $\times10^{-40}$  & 7.8 $\times10^{-48}$  & 1.5 $\times10^{-40}$  & 1.2 $\times10^{22}$  & 6.2 $\times10^{21}$   & $t\bar{t}$ (52\%), $W^+W^-$ (25\%), $ZZ$ (19\%), $Zh$ (3\%), $gg$ (1\%)                                                                                         & 79.1 \\
275.4  & 5.7 $\times10^{-27}$  & 0.116  & 3.5 $\times10^{-45}$  & 7.3 $\times10^{-41}$  & 3.4 $\times10^{-45}$  & 7.0 $\times10^{-41}$  & 2.6 $\times10^{21}$  & 1.3 $\times10^{21}$   & $W^+W^-$ (46\%), $ZZ$ (37\%), $b\bar{b}$ (12\%), $\tau^+\tau^-$ (3\%), $Zh$ (2\%),                                                                              & 95.6 \\  &&&&&&&&& $gg$ (1\%) & \\
799.4  & 2.9 $\times10^{-29}$  & 0.046  & 8.0 $\times10^{-47}$  & 1.4 $\times10^{-43}$  & 3.0 $\times10^{-47}$  & 5.2 $\times10^{-44}$  & 3.9 $\times10^{17}$  & 1.9 $\times10^{14}$   & $gg$ (64\%), $\tau^+\tau^-$ (18\%), $b\bar{b}$ (16\%), $\gamma\gamma$ (2\%)                                                                                & 0.0  \\
971.6  & 1.4 $\times10^{-26}$  & 0.100  & 5.2 $\times10^{-46}$  & 3.4 $\times10^{-43}$  & 4.4 $\times10^{-46}$  & 2.9 $\times10^{-43}$  & 2.5 $\times10^{18}$  & 1.1 $\times10^{18}$   & $W^+W^-$ (44\%), $ZZ$ (35\%), $W^\pm H^\mp$ (16\%), $A^0h$ (3\%), $Z\gamma$ (1\%)                                                                           & 5.2  \\
1019.1 & 4.3 $\times10^{-28}$  & 0.064  & 1.8 $\times10^{-47}$  & 1.7 $\times10^{-44}$  & 1.8 $\times10^{-47}$  & 1.7 $\times10^{-44}$  & 1.1 $\times10^{17}$  & 3.0 $\times10^{14}$   & $b\bar{b}$ (69\%), $t\bar{t}$ (17\%), $\tau^+\tau^-$ (13\%), $Zh$ (1\%)                                                                                      & 0.0  \\
1031.3 & 1.0 $\times10^{-26}$  & 0.106  & 9.7 $\times10^{-47}$  & 7.9 $\times10^{-44}$  & 8.6 $\times10^{-47}$  & 7.0 $\times10^{-44}$  & 4.7 $\times10^{17}$  & 1.0 $\times10^{17}$   & $W^+W^-$ (51\%), $ZZ$ (42\%), $W^\pm H^\mp$ (6\%), $A^0h$ (1\%), $Z\gamma$ (1\%)                                                                            & 0.0  \\
1078.2 & 1.0 $\times10^{-26}$  & 0.102  & 3.4 $\times10^{-45}$  & 1.4 $\times10^{-42}$  & 2.9 $\times10^{-45}$  & 1.2 $\times10^{-42}$  & 1.2 $\times10^{19}$  & 6.0 $\times10^{18}$   & $W^+W^-$ (51\%), $ZZ$ (37\%), $Zh$ (6\%), $t\bar{t}$ (4\%), $Z\gamma$ (1\%)                                                                                   & 13.5 \\
1157.6 & 1.2 $\times10^{-26}$  & 0.109  & 1.2 $\times10^{-45}$  & 2.5 $\times10^{-43}$  & 1.1 $\times10^{-45}$  & 2.2 $\times10^{-43}$  & 3.4 $\times10^{18}$  & 1.6 $\times10^{18}$   & $W^+W^-$ (35\%), $ZZ$ (29\%), $W^\pm H^\mp$ (20\%), $A^0h$ (5\%), $Zh$ (4\%),                                                                               & 6.5  \\  &&&&&&&&& $ZH$ (3\%), $A^0H$ (3\%), $t\bar{t}$ (1\%) & \\
1213.7 & 1.5 $\times10^{-26}$  & 0.103  & 8.7 $\times10^{-45}$  & 2.2 $\times10^{-42}$  & 7.5 $\times10^{-45}$  & 1.8 $\times10^{-42}$  & 2.2 $\times10^{19}$  & 1.1 $\times10^{19}$   & $W^+W^-$ (30\%), $W^\pm H^\mp$ (27\%), $ZZ$ (20\%), $A^0h$ (7\%), $t\bar{t}$ (5\%),                                                                           & 18.0 \\  &&&&&&&&& $Zh$ (4\%), $A^0H$ (3\%), $ZH$ (3\%), $b\bar{b}$ (1\%) & \\
\hline
\end{tabular}
\end{sidewaystable}

We also show a number of well-tempered neutralino benchmarks with faded symbols in Fig.\ \ref{fig::susy}, indicating that although they were consistent with all earlier data, they have since been excluded by LUX \cite{LUX13}.  One of these examples (the well-tempered neutralino MSSM-19 benchmark from Ref.\ \cite{SM13}) was already strongly excluded by the original LUX spin-independent limits.  The others satisfy the spin-independent limit, but are excluded by the recent \textsf{LUXCalc} \cite{LUXCalc} application of the LUX data to spin-dependent neutron scattering.\footnote{We note that shortly before this paper was accepted, LUX submitted their own official spin-dependent analysis \cite{LUX_SD_2016}, which improves on the \textsf{LUXCalc} limits.}  All of these models are strongly excluded by IceCube.

In Table\ \ref{bftable}, we give further details of all the benchmark models shown in Fig.\ \ref{fig::susy}.  These include cross-sections for annihilation and nuclear scattering ($\langle\sigma v\rangle$, $\sigma_{\rm SD}$, $\sigma_{\rm SI}$), relic densities ($\Omega h^2$), capture and annihilation rates ($C$, $A$), and dominant annihilation branching fractions (necessary to understand differences between the various well-tempered models).

The benchmark models we show here, whilst illustrative, are only isolated samples from the vast range of possible models in the MSSM.  A full statistical analysis of MSSM theories in the context of these data awaits their inclusion in large-scale global fits, as expected shortly from the GAMBIT Collaboration \cite{gambitweb}.

\section{Conclusions}
\label{conclusions}

We have presented a new analysis of data collected in the 79-string IceCube search for dark matter, taking into account energies of individual neutrino events.  This resulted in stronger spin-dependent limits on WIMP dark matter, in particular for high WIMP masses, and allowed us to rule out a number of MSSM models for the first time.  In the process, we developed an updated fast likelihood pipeline for event-level neutrino telescope DM search data, allowing it to be quickly and accurately applied to constrain essentially any dark matter model.  We have also provided a public code implementing the new likelihood (\textsf{nulike}), and made data from the 79-string IceCube DM search publicly available in a format compatible with its use.  Full details of the SUSY benchmarks and generic WIMP results presented in this paper are available as example programs in the public distribution of \textsf{nulike}.  Future improvements can be expected from applications of \textsf{nulike} to other models, and from the 86-string IceCube search for dark matter, which will include additional data and an improved energy proxy.

\acknowledgments{We acknowledge the support from the following agencies:
U.S. National Science Foundation-Office of Polar Programs,
U.S. National Science Foundation-Physics Division,
University of Wisconsin Alumni Research Foundation,
the Grid Laboratory Of Wisconsin (GLOW) grid infrastructure at the University of Wisconsin - Madison, the Open Science Grid (OSG) grid infrastructure;
U.S. Department of Energy, and National Energy Research Scientific Computing Center,
the Louisiana Optical Network Initiative (LONI) grid computing resources;
Natural Sciences and Engineering Research Council of Canada,
WestGrid and Compute/Calcul Canada;
Swedish Research Council,
Swedish Polar Research Secretariat,
Swedish National Infrastructure for Computing (SNIC),
and Knut and Alice Wallenberg Foundation, Sweden;
German Ministry for Education and Research (BMBF),
Deutsche Forschungsgemeinschaft (DFG),
Helmholtz Alliance for Astroparticle Physics (HAP),
Research Department of Plasmas with Complex Interactions (Bochum), Germany;
Fund for Scientific Research (FNRS-FWO),
FWO Odysseus programme,
Flanders Institute to encourage scientific and technological research in industry (IWT),
Belgian Federal Science Policy Office (Belspo);
University of Oxford, United Kingdom;
Marsden Fund, New Zealand;
Australian Research Council;
Japan Society for Promotion of Science (JSPS);
the Swiss National Science Foundation (SNSF), Switzerland;
National Research Foundation of Korea (NRF);
Danish National Research Foundation, Denmark (DNRF);
Science and Technology Facilities Council, United Kingdom (STFC).
We also thank the GAMBIT DM and Collider Workgroups for code testing of \textsf{nulike}.
}

\bibliography{DMbiblio,SUSYbiblio}

\end{document}